\documentclass[preprint,12pt,number]{elsarticle}

\usepackage[colorlinks=true, linkcolor=blue, urlcolor=blue]{hyperref}
\usepackage{amssymb}
\usepackage{amsmath}
\usepackage{svg}
\usepackage{mdframed}
\usepackage{tabularx}
\usepackage{lineno}
\usepackage{float}
\usepackage{upgreek}
\newcommand{\mum}{\,\upmu\mathrm{m}}
\usepackage{multicol}
\journal{Chemical Engineering Journal}
\usepackage[version=4]{mhchem}
\begin{document}

\begin{frontmatter}

\title{Comprehensive Study of 3D Liquid Flow Fields in Additive Manufactured Structures for SMART Reactors Using Large-Scale Vertical Magnetic Resonance Imaging and Computational Fluid Dynamics} 

\author[IMS]{Timo Merbach}
\ead{timo.merbach@tuhh.de}
\cortext[cor1]{Corresponding author}
\author[IPI]{Muhammad Adrian}
\author[IMS]{Christoph Wigger}
\author[IMS]{Selma Iraqi Houssaini}
\author[IMS]{Benedict Bayer}
\author[IBI,UKE]{Artyom Tsanda}
\author[ISM]{Serhan Acikgöz}
\author[IMS]{Christian Weiland}
\author[IMS]{Felix Kexel}
\author[ISM,FRA]{Dirk Herzog}
\author[IMS]{Marko Hoffmann}
\author[ISM,FRA]{Ingomar Kelbassa}
\author[IBI,UKE,FRALU]{Tobias Knopp}
\author[IPI]{Alexander Penn}
\author[IMS]{Michael Schlüter}

\affiliation[IMS]{organization={Institute of Multiphase Flows},
            addressline={Hamburg University of Technology, Eißendorfer Straße 38}, 
            city={21073 Hamburg},
            country={Germany}}

\affiliation[IPI]{organization={Institute of Process Imaging},
            addressline={Hamburg University of Technology, Denickestraße 17}, 
            city={21073 Hamburg},
            country={Germany}}

\affiliation[IBI]{organization={Institute for Biomedical Imaging},
            addressline={Hamburg University of Technology, Am Schwarzenberg-Campus 3}, 
            city={21073 Hamburg},      country={Germany}}

\affiliation[UKE]{organization={Section for Biomedical Imaging},
     addressline={University Medical Center Hamburg-Eppendorf, Lottestraße 55}, 
            city={22529 Hamburg},
            country={Germany}}

\affiliation[ISM]{organization={Institute for Industrialization of Smart Materials},
            addressline={Hamburg University of Technology, Harburger Schloßstraße 28}, 
            city={21079 Hamburg},
            country={Germany}}

\affiliation[FRA]{organization={Fraunhofer Research Institution for Additive Manufacturing Technologies IAPT},
            addressline={Am Schleusengraben 14}, 
            city={21029 Hamburg},
            country={Germany}}

\affiliation[FRALU]{organization= {Fraunhofer Research Institution for Individualized Medical Technology and Engineering IMTE},
            addressline = {Mönkhofer Weg 239a},
            city = {23562 Lübeck},
            country = {Germany}}

\begin{abstract}
Triply Periodic Minimal Surface~(TPMS) structures have emerged as a new class of porous materials with variable geometries and favourable transport properties, making them promising for reactor internals in chemical engineering. However, experimental data on internal TPMS flow behaviour are still limited. To address this gap, the flow behaviour in additively manufactured TPMS structures is analysed using three-dimensional Magnetic Resonance Imaging~(MRI) velocimetry in a large-bore vertical 3~T MRI system, in cylindrical columns of $38~\text{mm}$ diameter and Reynolds numbers between 50 and 300. Three different TPMS geometries are investigated, and consistency between Computational Fluid Dynamics~(CFD) simulations and experimentally measured MRI velocity fields is established through cross-validation. The MRI system provides fully three-dimensional velocity fields with a divergence deviation below $6~\%$. MRI revealed distinct flow features: the Gyroid~TPnS exhibited pronounced channelling, while the Schwarz-Diamond~TPSf showed merge-split behaviour, achieving a $46~\%$ increase in lateral mixing compared to the Gyroid TPnS structures. Numerical simulations reproduce the flow features and show agreement with the MRI data. The combined methodology demonstrates the suitability of MRI velocimetry for the experimental validation of CFD simulations and establishes a robust foundation for future studies of heat and mass transfer, as well as reactive flow, in structured reactor systems.
\end{abstract}


\end{frontmatter}

\section{Introduction}
\label{sec:Int}
Structured internals are widely applied in chemical engineering to enhance reactor performance compared with packed beds~\cite{cybulski2005structured, pangarkar2008structured, kapteijn2022structured, fratalocchi2022packed}. Their implementation, particularly in chemical and biochemical reaction systems with catalytically active surfaces, improves heat and mass transfer, increases process efficiency through higher volume-specific surface areas, and enhances product selectivity~\cite{pangarkar2008structured, feng2022triply, eckendorfer2024periodic}. \par 
While structured internals are designed to improve flow and transport properties, many industrially relevant chemical processes rely on fixed-bed reactors, which are characterised by randomly packed particles~\cite{bai2009coupled}. Such conventional reactors often face several operational limitations, including high pressure drops, dead zones, bypasses, uneven flow distribution, and consequently non-uniform access to the catalytic surface. Furthermore, the random packing of catalyst particles restricts scalability, while offering only limited flexibility in reactor design~\cite{cybulski2005structured}. To address these limitations, structured reactor concepts such as monolithic reactors are already in use. These systems consist of arrays of parallel channels with catalytically coated walls, providing low pressure drops and well-defined flow paths~\cite{cybulski2005structured}. However, the absence of cross-mixing between adjacent channels can limit their overall efficiency. Exchange between neighbouring channels can be achieved through so-called membrane catalysts, where transport predominantly occurs by diffusion, often resulting in mass-transfer limitations~\cite{cybulski2005structured, eckendorfer2024periodic}. \par
An emerging alternative to these conventional reactor designs is the implementation of geometries based on Triply Periodic Minimal Surfaces~(TPMS), which overcome the limitations of both randomly packed beds and monolithic reactors~\cite{baena2021stepping}. These analytically defined periodic implicit surfaces are generated from combinations of trigonometric functions and contain no sharp edges or junctions. As a consequence, they exhibit zero mean curvature and form smooth geometries~\cite{feng2022triply, feng2018porous}. Moreover, these structures offer high design flexibility, as their volume-specific surface area, porosity, and the geometry itself can be adjusted flexibly~\cite{feng2022triply}. TPMS-based architectures combine high mechanical stability with extensive interfacial area and favourable transport characteristics. Their interconnected channel networks promote homogeneous reactant distribution, enhanced mixing, and efficient heat and mass transfer~\cite{fratalocchi2022packed, feng2022triply,eckendorfer2024periodic,gado2024triply}. Recent advances in additive manufacturing have transformed the production of complex geometries through a layer-by-layer material deposition process~\cite{serhan2025}. This enables the manufacturing of TPMS structures with high precision and reproducibility from diverse materials such as metals or polymers, making them particularly advantageous for applications in reaction and heat transfer engineering~\cite{reynolds2025heat}. \par
The Collaborative Research Centre 1615 \textit{SMART Reactors}, funded by the German Research Foundation~(DFG), aims to develop novel reactor concepts capable of addressing the challenges of future chemical engineering applications. These reactor systems are engineered for sustainable and autonomous operation, exhibiting high adaptability to fluctuating process conditions due to the fluctuating quality of renewable reactants. Achieving such performance requires precise definition and control of the reactor geometry, since the internal structure fundamentally governs fluid dynamics, mixing behaviour, and mass transfer. Consequently, understanding the flow behaviour within the geometries of TPMS is essential for characterising flow patterns and local transport phenomena, which ultimately influence reaction yields. TPMS structures induce flow disturbances by geometry enhancing both heat and mass transport, making such structures particularly well-suited for applications in reactors~\cite{gado2024triply}. To characterise these flow patterns, suitable non-invasive measurement techniques are required. \par
Established non-invasive flow measurement techniques such as Particle Image Velocimetry~(PIV) and Particle Tracking Velocimetry~(PTV) require optical access to the flow domain. Although PIV has been successfully applied to refractive index-matched porous media~\cite{northrup1991fluorescent, sen2012optical}, and its application to TPMS structures has recently been demonstrated by Li~\textit{et al.}~\cite{li2024visual}, these studies remain restricted to thin sections of the structures. This limitation arises from refractive index variations induced by the complex geometry, which lead to severe image distortion and limits optical accessibility. \par
Tomographic methods overcome the limitations of optical techniques, as they do not require optical access. For studies of TPMS structures, spatial resolution is a critical factor, since industrially relevant designs with high volume-specific surface areas feature fluid channel dimensions on the order of a few millimetres. Resolving the internal flow structures therefore requires sub-millimetre spatial resolution. In this context, several tomographic approaches have been employed for the study of liquid flows in porous media~\cite{hampel2022review}. 
Electrical Capacitance Tomography~(ECT) and its advanced form, Electrical Capacitance Volume Tomography~(ECVT), are established techniques for flow imaging, with ECVT in particular having been applied to flow measurements in additively manufactured lattice structures~\cite{spille2021electrical}. However, both techniques do not provide sufficient spatial resolution to resolve the flow field within individual channels. Ultrasound tomography provides improved spatial resolution $(2~\text{mm})$ compared to ECT, and can in principle capture multiphase flow phenomena~\cite{hampel2022review}. However, the strong scattering of acoustic waves at solid-liquid interfaces, as well as the internal geometry of TPMS, reduces signal quality and image reconstruction accuracy. Other tomographic approaches, such as gamma-ray and X-ray computed tomography, have been applied to flow imaging in columns with structured internals and periodic open-cell structures~\cite{hampel2022review, wagner2016hydrodynamics}. However, their spatial resolution is typically limited to the millimetre range, which is insufficient for resolving the flow field within TPMS structures~\cite{hampel2022review, schubert2011advanced}. In contrast, Magnetic Resonance Imaging~(MRI) is an established method for flow measurements in chemical engineering~\cite{gladden2013recent, gladden2017magnetic}. MRI is based on the detection of the response of nuclear spins to magnetic fields and radiofrequency~(RF) pulses~\cite{haacke1999magnetic}. Using phase-contrast velocity encoding techniques, MRI enables the quantification of three-dimensional velocity fields with sub-millimetre spatial resolution, making it suitable for studying fluid flow in TPMS structures~\cite{hampel2022review}. Thus, numerous MRI studies have already been conducted to analyse the flow behaviour in porous media. For example, gas-liquid flows have been measured in packed beds and open-cell foams~\cite{sankey2009magnetic, sadeghi2020full} and liquid flow through porous filters and particle fixed-beds have also been studied by MRI~\cite{sederman1997magnetic, huang2017adapted}. Also, MRI has been successfully applied to flow characterisation of TPMS structures, primarily focusing on the Schwarz-Diamond geometry~\cite{clarke2021investigation, clarke2023characterization, clarke2025investigation}. By comparison, while the Gyroid is another widely studied TPMS geometry~\cite{gado2024triply}, its flow behaviour has not been investigated using MRI. \par 
In the present study, the flow behaviour of three different additively manufactured TPMS geometries is characterised in a column with a diameter of $38~\text{mm}$ using a large-bore vertical 3~T MRI system, with three-dimensional velocity fields acquired with Reynolds numbers between 50 and 300. Reynolds number are defined for porous media according to Eq.~\ref{eq:reynolds}. 
Unlike previous studies, the present work addresses both Schwarz-Diamond and Gyroid-based structures, thereby closing gaps in the literature and advancing the field. Furthermore, this study demonstrates the impact of unit cell geometry on the flow behaviour by rotating the unit cell in the Gyroid-based structure. Unit cell rotation is known to alter the mechanical properties of Gyroid structures~\cite{rezapourian2024effect}, but its effect on internal flow has not been studied. \par
Beyond the investigated geometries, this work advances previous studies by enabling measurements in TPMS structures with a larger column diameter of $38~\text{mm}$ while maintaining a small hydraulic diameter, the relevant characteristic length for complex geometries. The use of a large-bore MRI system makes such measurements feasible. Small hydraulic diameters correspond to a high volume-specific surface area, making them particularly relevant for chemical engineering applications. In previous MRI-based studies, measurements have been conducted in TPMS structures with hydraulic diameters of 2~mm and 7.2~mm and corresponding column diameters of 9~mm and 29.5~mm, resulting in ratios of column diameter to hydraulic diameter ranging from 4.1 to 4.5~\cite{clarke2021investigation, clarke2023characterization, clarke2025investigation}. In contrast, the present configuration achieves a ratio of up to 10.4, thereby extending the accessible scale for MRI-based flow characterisation in TPMS structures. Thus, this setup not only represents a 30~\% larger and more industrially representative geometry, but also contains a larger number of inner channels.
In addition to experimental MRI velocimetry, Computational Fluid Dynamics~(CFD) is also subject of research for studying flow behaviour through TPMS structures. However, CFD simulations require validation to ensure that the predicted flow fields represent real systems. Traditionally, these simulations are validated using integral quantities such as pressure drop, since direct measurements of three-dimensional velocity fields in complex porous geometries are difficult to obtain~\cite{kumar2014investigation, singh2022hydrodynamics, vhora2024cfd}. MRI allows macroscopic flow characterisation by experimentally measuring spatially resolved velocity fields. This enables cross-validation between both methods, whereby the CFD simulations are validated using MRI measurements. \par
The structure of the paper is as follows: Sec.~\ref{sec:Met} introduces the studied TPMS structures, experimental and computational setups used in the present work. Secs.~\ref{sec:struc_inte} and~\ref{sec:velodata} are dedicated to the validation of the MRI system through quantitative assessments, including structural integrity, mass flow rate and the divergence-free criterion. Additionally, Sec.~\ref{sec:flow_dynamics} presents an analysis that combines the qualitative identification of characteristic flow structures with vorticity as a quantitative flow metric. Finally, consistency between the CFD simulations and the MRI measurements is established through cross-validation in Sec.~\ref{sec:CFD_MRI}.

\section{Methods}
\label{sec:Met}
The selection and fabrication of the used TPMS is adressed in the following. Both, the experimental and the computational setups are presented. Further, the methods and principles concerning the fluid dynamics and the MRI-based flow measurements are investigated. 
\subsection{Selection and manufacturing of TPMS}
\label{sec:AMLS}
\begin{figure}[h!]
\centering
\includegraphics[width=0.7\linewidth]{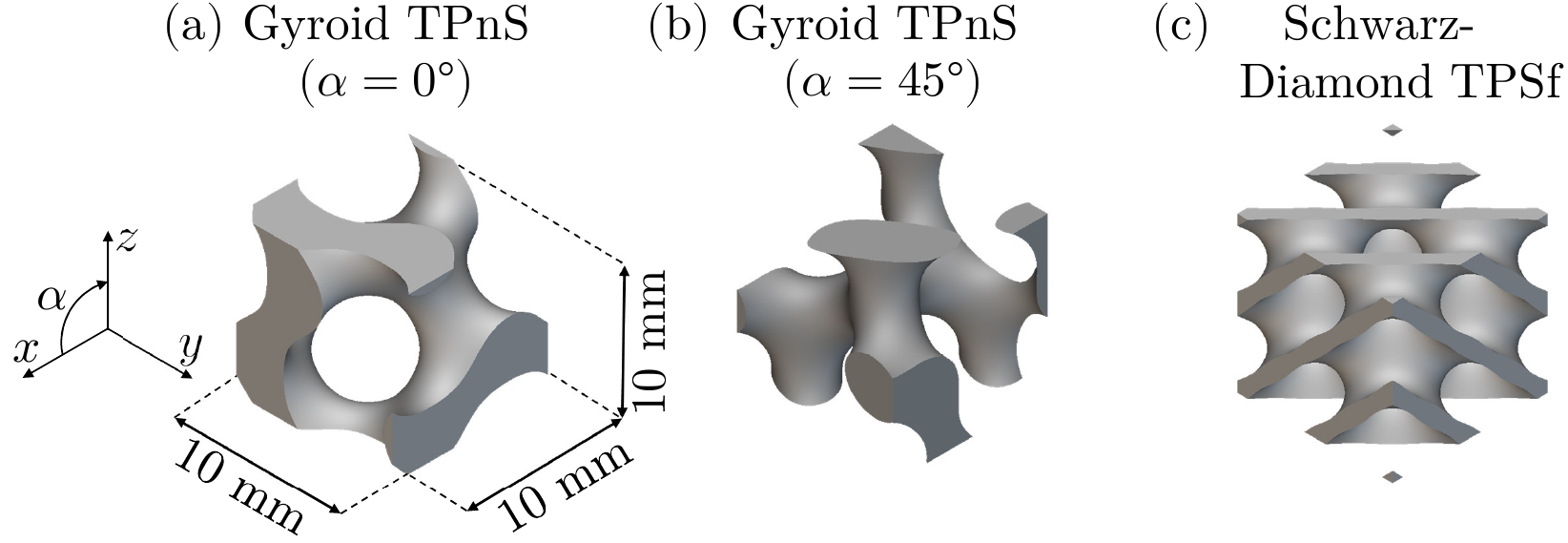}
\caption{Unit cells employed in this study: (a) Gyroid TPnS~$(\alpha = 0^\circ)$~(G), (b) Gyroid TPnS~$(\alpha = 45^\circ)$~(G45), and (c) Schwarz-Diamond TPSf~(SD). All structures exhibit a porosity of $\epsilon = 70\%$ and a unit cell size of $10 \times 10 \times 10~\text{mm}^3$.}\label{fig_unitcell}
\end{figure}

In this study, three distinct TPMS structures are examined as porous internals for chemical engineering applications. Two unit cells, a Gyroid Triply Periodic endo-Skeleton (TPnS) and a Schwarz-Diamond Triply Periodic Surface (TPSf), are generated. In addition to the standard Gyroid TPnS, a Gyroid TPnS geometry rotated by $45^\circ$ (see Fig.~\ref{fig_unitcell} for the definition of the angle orientation) is also examined, as this configuration is expected to exhibit improved flow characteristics, as discussed in the following. Figure~\ref{fig_unitcell} illustrates the unit cells, which are named following the nomenclature of Fisher \textit{et al.}~\cite{fisher2023catalog}. \par
After selecting the type of unit cell, each TPMS is characterised by three key parameters: the sheet thickness, the porosity, and the unit cell size, defined as the edge length of a three-dimensional cube. In this paper, all structures exhibit a unit cell size of $10 \times 10 \times 10~\text{mm}^3$ and a porosity of $\epsilon = 70~\%$, which can be chosen independently of one another. Consequently, the sheet thickness follows as the dependent variable of the other two. 

The two types of TPMS structures studied in this work differ primarily in their number of nodes and channels, as well as in their volume-specific surface areas. Both structures exhibit high volume-specific surface areas~$a$, with $a_\text{G} = 320~\text{m}^2\ \text{m}^{-3}$, $a_\text{G45} = 322~\text{m}^2\ \text{m}^{-3}$ and $a_\text{SD} = 767~\text{m}^2\ \text{m}^{-3}$, making them particularly suitable for heterogeneous catalysis and heat-transfer applications. Due to its large surface area, the Schwarz-Diamond geometry has already been employed in studies of heterogeneous catalysis~\cite{lei2019feasibility}. Previous studies have also shown that the Gyroid-based structure outperforms conventional monolithic designs in terms of yield and mass-transfer performance, resulting in higher overall efficiency~\cite{baena2021stepping}. Moreover, the Gyroid geometry achieves superior heat-transfer rates under turbulent flow conditions, whereas the Schwarz-Diamond exhibits excellent thermal performance even under laminar conditions~\cite{yeranee2022review}.
These favourable properties arise not only from the high volume-specific surface areas but also from the tortuous geometry of the structures. Both Schwarz-Diamond and Gyroid structures exhibit intensified mixing and flow homogenisation~\cite{gado2024triply}. The geometric configuration of the Gyroid structure generates helical flow patterns in both clockwise and counter-clockwise directions, while the Schwarz-Diamond TPSf additionally produces characteristic merge-split flow patterns~\cite{gado2024triply}. Although both structures exhibit pressure drops of similar magnitude at low Reynolds numbers~\cite{gado2024triply, iyer2022heat}, the Gyroid TPnS inherently induces channelling effects due to its continuous pathways aligned with the main flow direction~\cite{padrao2024new}. To counteract this tendency and enhance lateral mixing, a Gyroid TPnS rotated by $\alpha = 45^\circ$ around the $y$-axis is therefore included in the analysis (see Fig.~\ref{fig_unitcell}(b)). As stated in Sec.~\ref{sec:Int}, the impact of unit cell rotation on internal flow behaviour has not yet been studied and is addressed in the present study. To investigate such geometric effects, the TPMS structures considered in this work are constructed by periodic replication of a single unit cell, as described below. \clearpage

\begin{figure*}[h!]
\centering
\includegraphics[width=1.0\linewidth]{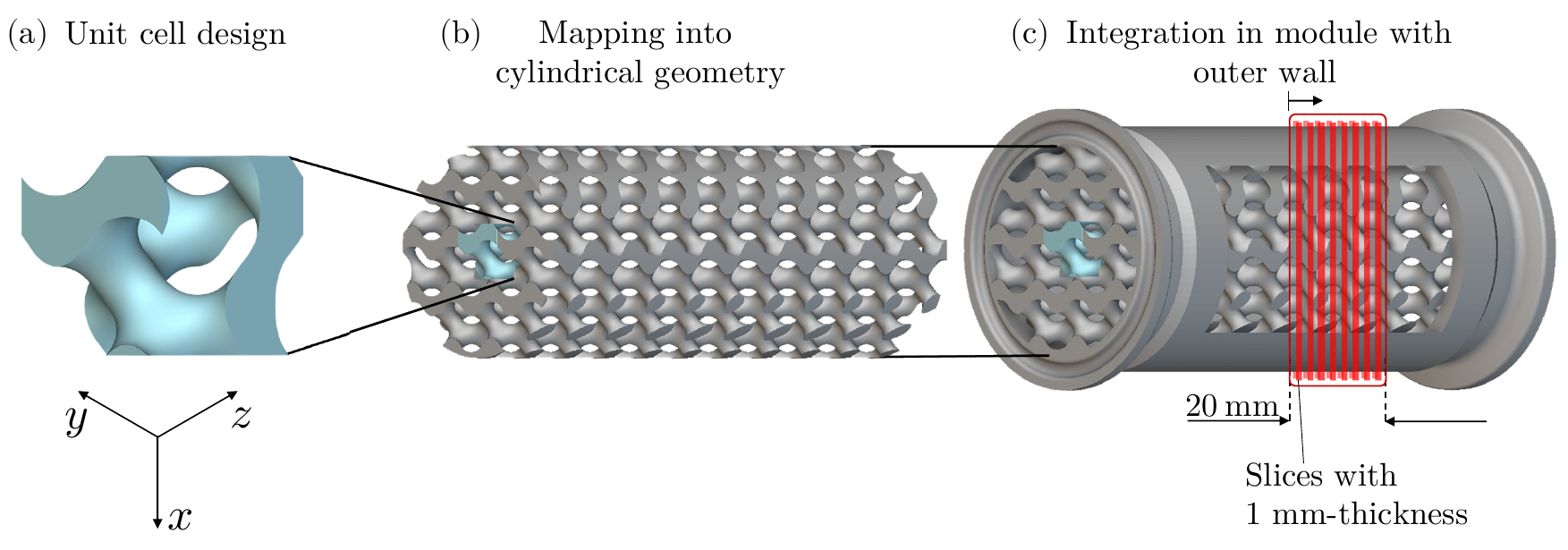}
\caption{Generation process of the structures, exemplified by the Gyroid TPnS~$(\alpha = 0^\circ)$. (a)~Unit cell design. (b)~Assembly of the structure as a rectangular block and subsequent mapping into a cylindrical geometry. (c)~Integration of the surrounding wall and clamping connectors into the cylindrical design. The red section indicates the FOV for the MRI measurements, covering a length of 20~mm with 20 slices of 1~mm thickness. For clarity, the figure shows a reduced number of slices.}\label{fig:module_gen}
\end{figure*}

Starting from the individual unit cell, the complete structures for the study are generated. Fig.~\ref{fig:module_gen} illustrates the generation process shown here using Gyroid TPnS~$(\alpha = 0^\circ)$ as an example. The structures are designed using nTopology~(nTopology Inc., USA), beginning with the individual unit cell design (see Fig.~\ref{fig:module_gen}(a)). Subsequently, rectangular blocks are generated and cut into cylindrical geometries with an inner diameter of $D = 38~\text{mm}$ (DN~40 in accordance with DIN~11866~\cite{DIN11866}) and a length of $L_\text{M} = 100~\text{mm}$ (see Fig.~\ref{fig:module_gen}(b)). These cylindrical structures are then integrated with a surrounding wall and clamping connectors in accordance with DIN~32676~\cite{DIN32676} (see Fig.~\ref{fig:module_gen}(c)). In addition, the Field Of View~(FOV) for the MRI measurements is illustrated, with further details provided in Secs.~\ref{sec:Exsetup} and~\ref{sec:MRI}. After the design process, the entire structure is fabricated in a single step to eliminate the channelling effects between the structure and the surrounding wall. Moreover, this approach represents real reactor configurations, as the integrated wall enhances heat transfer under both heating and cooling conditions~\cite{fratalocchi2022packed, serhan2025}.

The Gyroid TPnS~$(\alpha = 0^\circ)$ structure is fabricated by Laser-Based Powder Bed Fusion of Polymers~(PBF-LB/P), often also referred to as selective laser sintering, using an EOS Eosint P396 system (\textit{EOS GmbH, Germany}). In this process, a laser selectively melts layers of polyamide 12 powder $($median particle diameter of $d_{50} = 51-61~\mum)$ to form the final structure. After fabrication, the part is cleaned by sandblasting to remove residual powder. \par 
However, a limitation of this technique is the removal of unsintered powder from geometries without continuous channels and limited accessibility due to the surrounding wall. As a result, complete powder removal could not be achieved for the Schwarz-Diamond TPSf and rotated Gyroid TPnS structures following the PBF-LB/P process. Therefore, these two structures are fabricated using a Formlabs Form 3+ system (\textit{Formlabs, USA}), which employs Vat Photopolymerisation, curing by Ultra Violet Laser beam exposure~(VPP-UVL). In this process, often also referred to as stereolithography~(SLA), a photosensitive polymer (Clear V4 resin (\textit{Formlabs, USA})) is cured layer by layer to build the TPMS modules. The VPP-UVL printer provides a spatial resolution of $0.25~\text{mm}$. \par 
The main distinction between the two manufacturing methods lies in the surface quality. Since PBF-LB/P relies on a powder-based process, the resulting surface is rougher compared to the smooth surface obtained via VPP-UVL. However, the influence of surface roughness on the macroscopic flow behaviour is expected to be negligible. The roughness, which is expected to correspond approximately to the median particle diameter of the powder, remains well below the selected spatial resolution of the MRI (see Sec.~\ref{sec:MRI}) and therefore does not introduce measurement bias on the scale relevant to this study. Consequently, both materials remain fully suitable for MRI-based flow analysis, and the comparison with CFD is not affected.

\subsection{Experimental setup and fluid system}
\label{sec:Exsetup}
\begin{figure}[ht]
\centering
\includegraphics[width=0.98\linewidth]{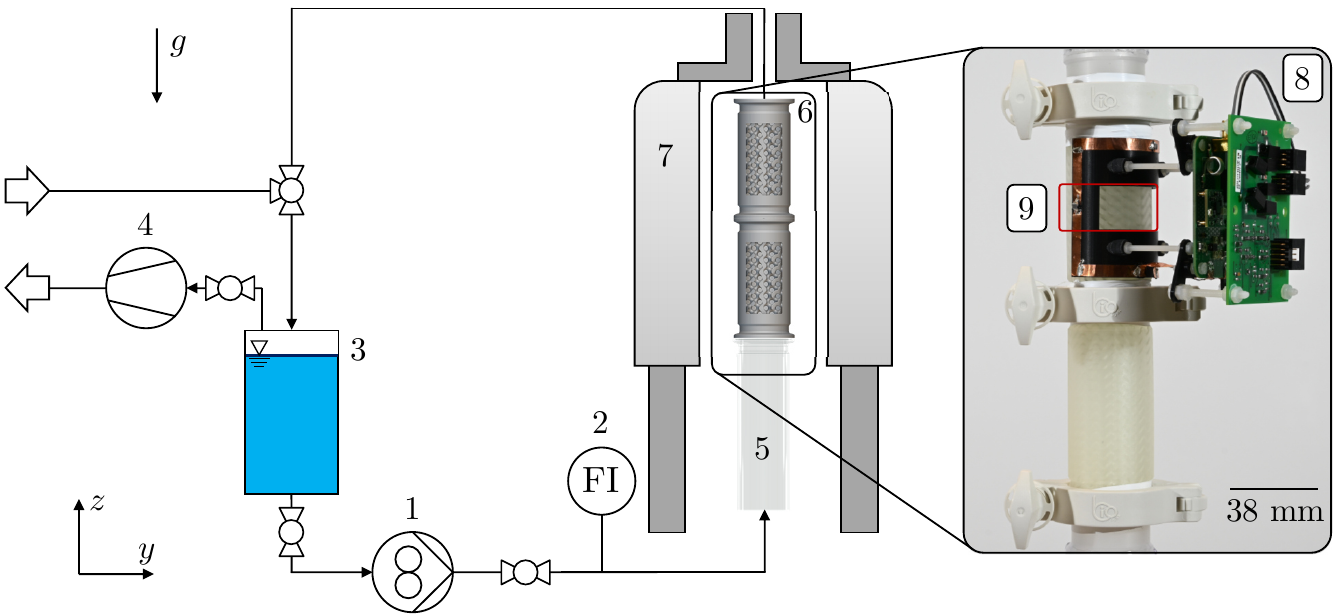}
\caption{Left: Schematic flow diagram of the experimental setup showing the gear pump~(1), the Coriolis MFM~(2) for monitoring the mass flow rate, the degassing vessel~(3) with the vacuum pump~(4), the inlet acrylic pipe~(5), the TPMS modules~(6), and the MRI system~(7). The main flow direction through the TPMS modules is oriented opposite to the gravitational acceleration~$g$. Right: Close-up of the analysed area is illustrated, including the custom-built RF receive coil with single-channel~(8) and the FOV within the module~(9).}\label{fig:flow-shart}
\end{figure}

The experimental setup to investigate the flow behaviour within the TPMS structures is shown in Fig.~\ref{fig:flow-shart}. It consists of a gear pump~(1) (\textit{Gather Industries, Germany}), a Coriolis Mass Flow Meter~(MFM, 2) (\textit{Endress \& Hauser, Germany}) for monitoring the mass flow rate. According to the manufacturer, the measurement deviation of the MFM is less than $\pm 1.25~\%$ of the measured mass flow rate~\cite{Endress}. A degassing vessel~(3) (\textit{Binder GmbH, Germany}) is connected to a vacuum pump~(4) (\textit{Edwards Vacuum, UK)}, and an acrylic pipe~(5) is installed, followed by the TPMS modules~(6) analysed in this study, as introduced in Sec.~\ref{sec:AMLS}. The upstream acrylic pipe has an inner diameter of $D = 38~\text{mm}$ (DN~40) and a length of $L_\text{E} = 760~\text{mm}$. This length helps to minimise the influence of upstream flow development and capture intrinsic flow characteristics within the TPMS structures. Two TPMS modules, each $L_\text{M} = 100~\text{mm}$ in length, are mounted downstream of the acrylic pipe and connected using nylon BioClamps (\textit{BioPure Technology Ltd, UK}) as clamping connections. \par

The acrylic pipe together with the mounted structures is positioned inside the MRI system~(7) for analysis. Moreover, Fig.~\ref{fig:flow-shart} provides a close-up view of the experimental setup with the RF receive coil~(8), which has a clip-on feature for easy switching between different structures. The FOV~(9) for the MRI measurements is highlighted in accordance with Fig.~\ref{fig:module_gen}. Measurements are taken in the centre of the second module, approximately after a length of $L_\text{E, M}=150~\text{mm}$ downstream, where sufficient momentum exchange with the structure ensures that the flow conditions are independent of the inlet effects. This observation is consistent with the findings of Hawken \textit{et al.}, indicating that entrance effects are negligible for $L_\text{E, M}>11D_\text{h}$, where $D_\text{h}$ denotes the hydraulic diameter defined in Eq.~\ref{eq:Dh}~\cite{hawken2023characterization}.

The working fluid is deionised water supplemented with copper sulphate pentahydrate (\textit{Carl Roth GmbH + Co. KG, Germany}, CAS: 7759-99-8, $\text{purity} \geq 98~\%$) at a concentration of $c_{\ce{CuSO4}}=34~\text{mmol L}^{-1}$, which has beneficial effects for MRI measurements, as discussed in Sec.~\ref{sec:MRI}. The used fluid is maintained at a temperature of $T = (25 \pm 1)~^\circ\text{C}$ with a density of $\rho = (1002.9 \pm 0.3)~\text{kg m}^{-3}$, interpolated from \cite{novotny1988densities, stephan2019vdi}, and a viscosity of $\eta = (0.890 \pm 0.020)~\text{mPa s}$~\cite{stephan2019vdi}. For the viscosity, the value of pure water is used, as a minor addition of copper sulphate is estimated to cause a deviation of less than 2~\%~\cite{stephan2019vdi, motin2004temperature}. The standard deviation is indicated by $\pm$ throughout this paper. \par 
The working fluid is pumped from the degassing vessel through the gear pump and then through the TPMS modules. Before each experiment, it is essential to ensure that no gas bubbles remain within the structures, as they may distort the flow field by blocking channels or adhering to internal surfaces. The narrow channels and tortuous pathways of the TPMS geometries promote bubble accumulation, and therefore a dedicated procedure is applied to minimise bubble entrapment. For each experiment, the degassing vessel is filled with the working fluid, sealed, and evacuated using the vacuum pump until an absolute pressure of $p < 20~\text{mbar}$ is reached and maintained for at least $20~\text{minutes}$. This step removes dissolved gases that would otherwise emerge within the structures due to pressure drops during operation. The vessel is subsequently returned to atmospheric pressure $(p = 1.013~\text{bar})$, and the valves are adjusted to establish a closed-loop circulation of the working fluid. The remaining bubbles are eliminated by repeated cycles of oscillatory volume fluxes. The degassing procedure is repeated until no bubbles are visible in the MRI images and no gas release is observed after the modules. Once a bubble free state is reached, MRI measurements are initiated. \par 
\begin{table*}[h!]%
\centering
\caption{Steady-state operating conditions, including mass flow rate $\dot{M}$, superficial velocity $w_\text{S}$, and Reynolds number for porous media~$Re_\text{S}$ for the three operating points.}
\vspace{10pt}
\begin{tabular}{l c c c}
\hline Mass flow rate $\dot M\ /\ \text{kg min}^{-1}$ & $0.755 \pm 0.03$ & $1.15 \pm 0.01$ & $1.50 \pm 0.01$ \\ 
  Superficial velocity $w_\text{S}\ /\ \text{mm s}^{-1}$ & 11.1 & 16.9 & 22.0 \\ \hline
   \multicolumn{4}{c}{Reynolds number for porous media~$ Re_\text{S}\ /\ -$} \\ \hline
  $\text{Gyroid TPnS}\ (\alpha = 0^\circ$)~(G) & 156 & 237 & 310 \\
  $\text{Gyroid TPnS}\ (\alpha = 45^\circ$)~(G45) & 155 & 236 & 308 \\
  $\text{Schwarz-Diamond TPSf}$~(SD) & 65.0 & 99.1 & 129 \\ \hline
\end{tabular}
\label{tab:operating_cond}
\end{table*}
Measurements are performed under different operating conditions that are characterised by varying superficial velocities. Tab.~\ref{tab:operating_cond} summarises the operating conditions, including the measured mass flow rate~$\dot M$, the superficial velocity~$w_\text{S}$ in the empty pipe, and the Reynolds number within the structures~$Re_\text{S}$ (see Eq.~\ref{eq:reynolds}). In porous structures, the Reynolds number must be defined using an appropriate characteristic length. Various definitions have been proposed, as reviewed by Eckendörfer \textit{et al.}~\cite{eckendorfer2024periodic}, with the hydraulic diameter most commonly used for $Re_\text{S}$ in porous or structured media. The definition of the hydraulic diameter~$D_\text{h}$, which is given by 
\begin{equation}
\label{eq:Dh}
D_\text{h} = 4 \epsilon \frac{V_\text{W}}{A_\text{W,\ S}} = \frac{4\epsilon}{a}\; ,
\end{equation}
where $V_\text{W}$ denotes the wetted (free) volume and $A_\text{W,\ S}$ the wetted surface area, $\epsilon$ the porosity and $a$ the volume-specific surface area~\cite{dietrich2009pressure}. For the selected structures, the corresponding geometric properties are determined using nTopology, yielding in hydraulic diameters of $D_\text{h, G} = 8.8~\text{mm}$ for the Gyroid TPnS, $D_\text{h, G45} = 8.7~\text{mm}$ for the rotated Gyroid TPnS, and $D_\text{h, SD} = 3.7~\text{mm}$ for the Schwarz-Diamond TPSf structure. On this basis, the Reynolds number for porous media~$Re_\text{S}$ can be expressed as
\begin{equation}
\label{eq:reynolds}
Re_\text{S} = \frac{\rho w_\text{S} D_\text{h}}{\eta \epsilon}\; ,
\end{equation}
with $\rho$ corresponding to the fluid density, $\eta$ to the dynamic viscosity of the fluid, and $w_\text{S}$ to the superficial velocity~\cite{dietrich2009pressure}. For all operating conditions studied in the Schwarz-Diamond TPSf geometry (see Tab.~\ref{tab:operating_cond}), the Reynolds number ranges from $10 \lesssim Re_\text{S} \lesssim 150$, corresponding to the Darcy-Forchheimer regime, in which (non-linear) laminar flow prevails~\cite{sadeghi2020full, dybbs1984new}. In contrast, the Reynolds numbers for the Gyroid TPnS structures fall within $150 \lesssim Re_\text{S} \lesssim 300$, corresponding to the post-Forchheimer regime in which unsteady flows with laminar wake oscillations occur. Above $Re_\text{S} \approx 250$, vortex formation occurs, and for $Re_\text{S} \gtrsim 300$, the flow becomes turbulent, exhibiting unsteady and chaotic behaviour~\cite{dybbs1984new}. The structures are primarily analysed and compared within the Darcy-Forchheimer regime, where the flow remains laminar.

\subsection{MRI velocimetry}
\label{sec:MRI}
MRI velocimetry measurements are performed on a worldwide unique large-bore 3~T vertical MRI system with an inner diameter of 400~mm located at Hamburg University of Technology. The vertical configuration of the MRI offers several experimental advantages: it allows the study of longer setups with larger diameters, while maintaining a flow orientation consistent with most process equipment such as plug flow reactors or distillation columns. Moreover, gas bubbles can rise naturally, preventing their accumulation in the measurement section and avoiding artefacts in the recorded velocity fields. \par 
The MRI system is equipped with a birdcage RF coil for homogeneous signal excitation and a custom-built single-channel RF receive coil for signal detection. Measurements are conducted in the FOV inside of the structures, which is illustrated in Figs.~\ref{fig:module_gen}(c) and~\ref{fig:flow-shart}. The setup is placed so that the FOV is positioned at the isocentre of the scanner to ensure maximal magnetic field homogeneity. A $34~\text{mmol L}^{-1}$~copper sulphate solution (see Sec.~\ref{sec:Exsetup}) is used as the working fluid due to its relaxation properties for MRI-based flow measurements~\cite{que2009copper}. At the used copper sulfate concentration, the solution yields a short longitudinal (spin-lattice) relaxation time of $T_1 = 20~\text{ms}$~\cite{callaghan2011translational, d2017effect}, measured using a 60~MHz benchtop nuclear magnetic resonance spectrometer (\textit{Spinsolve~60, Magritek, Germany}). The short relaxation time~$T_1$ enables fast re-excitation of the spin system, which is essential for time-efficient MRI velocimetry~\cite{fram1987rapid}. \par
MRI-derived velocity fields are quantified by employing bipolar flow-encoding gradients, which encode coherent spin motion as velocity-dependent phase shifts in the MR signal~\cite{clarke2021investigation}. Following correction for background phase contributions, the residual phase is directly proportional to the local fluid velocity and is used to compute spatially-resolved velocity fields~\cite{o1985nmr}. Measurements are conducted using a gradient echo pulse sequence with a Multi-two-Dimensional~(M2D) acquisition method, enabling three-dimensional reconstruction of the velocity fields within the TPMS structures. The measurement parameters employ a flip angle of $90^\circ$, an echo time of 6.8~ms and a repetition time of 150~ms. The measurements are acquired with a spatial resolution of $0.25 \times 0.25~\text{mm}^2$ in the lateral plane, corresponding to a pixel-to-millimetre ratio of $4.4~\text{px mm}^{-1}$. The FOV in the lateral plane is $40 \times 40~\text{mm}^2$, while the slice thickness is $1~\text{mm}$ per slice. Consecutive slices are recorded along the axial direction, covering a total distance equivalent to twice the unit cell length (20~mm) (see Fig.~\ref{fig:module_gen}(c)). The total acquisition time for the measurement per operating flow rate condition is approximately 32~minutes. In this study, a high spatial resolution of $250~\mum$ in lateral direction is chosen to resolve the channels inside the TPMS structures. In this work, the achievable temporal resolution restricts the analysis to steady-state flow conditions, which is assumed for the selected Reynolds numbers.
\par

\subsection{Fluid-dynamic fundamentals}
\label{sec:fluid}
MRI-based flow measurement is a well-established technique. However, validation is essential in MRI velocimetry because measurement uncertainty and biases can arise from imposed assumptions, signal noise, and imperfections in the experimental setup. To ensure the reliability of the results, the validation of the measured velocity fields is performed using fundamental fluid-dynamic principles. From the three-dimensional velocity field, the mass flow rate~$\dot M$ can be determined and directly compared with the measured value from the MFM. The mass flow rate is obtained from the data by multiplying the fluid density by the axial velocity~$w(x,y)$ integrated over the free cross-sectional area~$A_\text{S}$~\cite{clarke2021investigation, spurk2007fluid}. In the case of discrete data, the axial velocity of each pixel~$w_i(x,y)$ is multiplied by the area of pixel~$A_\text{P} = 0.0517~\text{mm}^2\ \text{px}^{-1}$ and the fluid density. The sum of all pixels~$n$ yields in

\begin{equation}
\label{eq:mass-flow}
\dot M = \rho \iint_{A_\text{S}} w(x,y)\text{d}A \approx \rho \sum_{i = 1}^n \left ( w_i(x,y) \right ) A_\text{P}.
\end{equation}

Since the mass flow rate validates only for the axial component of the velocity, while MRI velocimetry and CFD provide all three velocity components, the divergence is additionally assessed. For incompressible fluids, the velocity field must satisfy the divergence-free condition for the velocity field~$\boldsymbol{u}$
\begin{equation}
\label{eq:div}
\text{div\ } \boldsymbol{u} = \frac{\partial u}{\partial x} + \frac{\partial v}{\partial y} + \frac{\partial w}{\partial z} = 0\; ,
\end{equation}
where $u$, $v$, and $w$ represent the velocity components in the $x$-, $y$-, and $z$-directions, respectively~\cite{spurk2007fluid}. Both Eqs.~\ref{eq:mass-flow} and~\ref{eq:div} are used to assess the physical consistency of the MRI-derived velocity data and thus serve as validation criteria. \par
Furthermore, the vorticity~$\omega_z$, defined as
\begin{equation}
\omega_z = \frac{\partial v}{\partial x} - \frac{\partial u}{\partial y}\; ,
\label{eq:vor}
\end{equation}
is evaluated~\cite{spurk2007fluid}. It serves as a key measure of rotational flow structures. Due to the discrete nature of the data, spatial derivatives are evaluated using finite-difference schemes, with second-order central differences applied in the interior and first-order one-sided differences at the boundaries. These operations are implemented in MATLAB (\textit{MathWorks Inc., USA}) using the built-in \texttt{gradient} function to estimate the spatial gradients. In addition to the local flow analysis based on differential operators, the hydraulic tortuosity~$T_\text{h}$ is introduced as a macroscopic quantity. It is defined as 
\begin{equation}
\label{eq:TH}
T_\text{h}(z) = \frac {\sum_{i=1}^n |\boldsymbol{u}_i(x,y,z)|}{\sum_{i=1}^n w_i(x,y,z)} \;,
\end{equation}
\noindent
where $|\boldsymbol{u}_i(x,y,z)|$ is the magnitude of the velocity vector and $w_i(x,y,z)$ is the axial velocity at the lateral position of each in-plane position~$x,y$~\cite{matyka2012calculate}. $n$ denotes the number of pixels, and $z$ corresponds to the slice location. The hydraulic tortuosity serves as a measure of the deviation of the flow from the axial direction and, consequently, characterises the degree of lateral mixing within the structure~\cite{clarke2021investigation, matyka2012calculate}.

\subsection{Computational setup}
\label{sec:Comsetup}
\begin{figure*}[h!]
\centering
\includegraphics[width=1.0\linewidth]{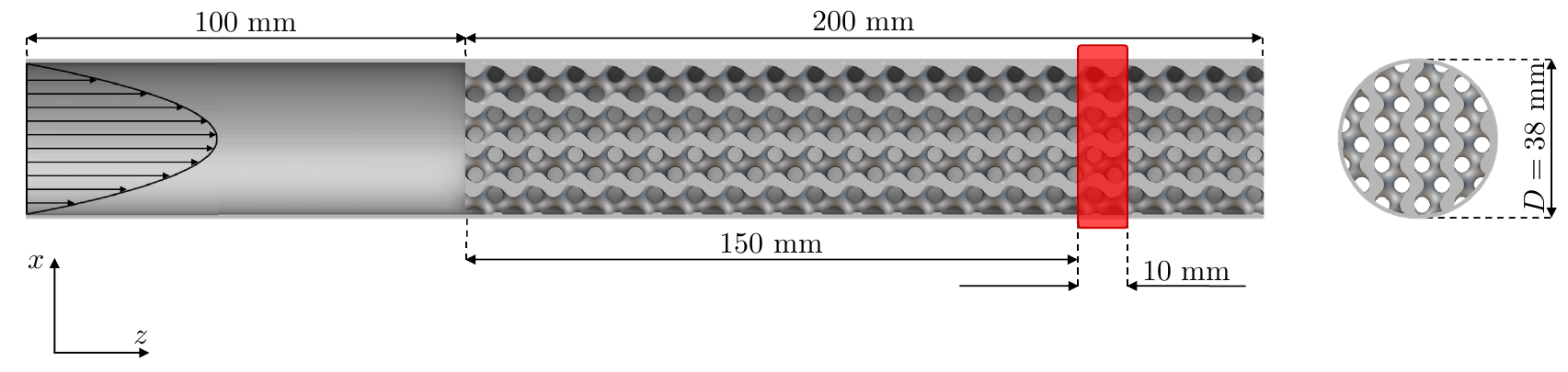}
\caption{Full-scale CFD configuration with an established flow profile, highlighting the analysed FOV (red), shown for the Gyroid TPnS $(\alpha = 0^\circ)$ structure.}\label{fig:comDomain}
\end{figure*}

The numerical flow simulations are conducted using the same Computer-Aided Design~(CAD) models that served as the basis for additive manufacturing. Thus, the computational domain comprises two cylindrical elements shown in Fig.~\ref{fig:module_gen}(c), supplemented by an additional pipe section to ensure the development of an established flow profile~(see Fig.~\ref{fig:comDomain}). Moreover, the simulations are performed under the same operating conditions as the experiments. \par 
All numerical simulations are performed using the open-source software \textit{OpenFOAM} (Version~2406) with the \textit{simpleFoam} solver, steady state flow conditions as indicated by the MRI measurement. Inlet velocity and pressure outlet boundary conditions are applied in addition to standard wall boundary conditions at the acrylic pipe and the TPMS modules, including a no-slip condition. The flow field is initialised with a parabolic velocity profile corresponding to the respective operating point. The initial pressure field is set to a uniform 0~Pa gauge pressure, as for incompressible media, the pressure levels are relative to a reference pressure. During initialisation first-order Gauss upwind schemes are employed for gradient and divergence terms, whereas for an accurate solution spatial discretisation is performed using the least squares method for gradients, Gauss QUICK schemes for divergence terms, and a second-order linear corrected scheme for Laplacian terms. Comprehensive information concerning the initial conditions, boundary conditions, numerical discretisation schemes, and solver settings are available in the associated research data~\cite{merbach2026data}. \par
Unstructured hexahedral meshes are generated using \textit{snappyHexMesh}. Four progressively refined grids are prepared to assess the reliability of the simulation results with varying grid resolution
\begin{equation}
    R^{*} = \frac{D}{\mathrm{max}(\Delta x, \Delta y, \Delta z)} \;,
\end{equation}
where $D=38~\mathrm{mm}$ corresponds to the characteristic length of the acrylic pipe (inner diameter) and $\Delta x$, $\Delta y$ and $\Delta z$ are the distance between nodes of the equidistant hexahedral mesh. The convergence study has been conducted for the Gyroid TPnS~$(\alpha=0^\circ)$ with grid resolution $R^{*} \in \{20,30,40,50\}$~\cite{merbach2026data}. To determine which grid resolution is sufficiently fine enough to deliver reliable and grid independent results, it is verified by the Grid Convergence Index~(GCI) proposed by Roache~\cite{roache1997quantification}. The GCI serves as an indicator for assessing the optimal computational grid by estimating the discretisation error based on Richardson's extrapolation~\cite{richardson1927viii}. Since the pressure inherently depends on the overall momentum balance and energy dissipation of the flow, it is highly sensitive to the accuracy of the resolved velocity fields that characterise the fluid dynamic state, thereby making it a suitable parameter for assessing the GCI. The refinement factor~$r$ is computed for a set of simulation, consisting of a coarse, medium, and fine grid. The order of discretisation is determined iteratively for a set of grids to account for potential non-monotone convergence. For a set of three simulations, a fine GCI
\begin{equation}
    \mathrm{GCI_{fine}^{12}} = F_\text{S} \cdot \frac{\varepsilon_{12}}{1-r^P}\;,
\end{equation}
and a coarse GCI
\begin{equation}
    \mathrm{GCI_{coarse}^{23}} = F_\text{S}\cdot \frac{\varepsilon_{23}}{1-r^P}\;
\end{equation}
can be approximated by the Richardson error estimator with the convergence index~$P$, the change of the numerical solution $\varepsilon$ and a safety factor $F_\text{S}=1.25$. For grid studies considering three or more grid solution, this safety is adequately sufficient~\cite{roache1997quantification}.
As shown in Tab.~\ref{tab:GCI}, the mesh configuration $R^{*}=50$ exhibits an exceptionally low grid convergence index $\mathrm{GCI_{fine}^{23}}=2.721 \cdot 10^{-6}$, indicating that further grid refinement is not necessary. Therefore, a grid configuration $(R^*=50)$ is used for this study. \par 
\begin{table*}[h]%
\centering
\caption{Grid convergence index analysis for different mesh types, showing average mesh size, resulting pressure drop, and $\text{GCI}_\text{fine}$}
\vspace{10pt}
\begin{tabular}{c c c c c}
  \hline 
 \parbox[c]{2.5cm}{Grid reso-\\ lution~$R^*$ / --} & \parbox[c]{2.5cm}{Average mesh \\ size~$h$ / $\mum$} & \parbox[c]{2.5cm}{Pressure drop\\$\Delta p$ / Pa} & $\mathrm{GCI_{course}^{12}}$ & $\mathrm{GCI_{fine}^{23}}$ \\ \hline
  20 & 279.1 & 5.722 & $3.520\cdot10^{-1}$ & --\\ 
  30 & 214.8 & 5.791 & $1.351\cdot10^{-4}$ & -- \\
  40 & 178.4 & 5.809 & -- & $1.630\cdot10^{-1}$\\
  50 & 154.0 & 5.810 & -- & $2.721\cdot10^{-6}$\\ \hline
\end{tabular}
\label{tab:GCI}
\end{table*}
The numerical data are used for cross-validation with the MRI measurements, whereby the CFD simulations are validated using the experimentally obtained velocity fields. MRI data are aligned with the CAD geometry to determine the axial position within the TPMS structure, as well as the orientation of the measured slice. The MRI data are rotated and shifted laterally to match the orientation of the CFD dataset. MRI data have an axial spatial resolution of 1~mm (see Sec.~\ref{sec:MRI}), whereas the CFD fields are generated of infinitesimal small slices generated at a spacing of $200~\mum$. Thus, to allow direct comparison, the CFD data must be volumetrically averaged over the corresponding 1~mm slice thickness. Without this averaging, the geometric variations of the structure along the axial direction are not properly represented. To align the CFD and MRI datasets, a matching procedure is applied in which the mean deviation between the first MRI slice and the corresponding CFD slice is evaluated. The axial position of the CFD data is adjusted iteratively until the smallest mean deviation with the corresponding MRI slice is achieved. The subsequent slice pairs are assigned at 1~mm intervals along the axial direction. Starting from this position, 1~mm-thick CFD slices are subsequently extracted for cross-validation analysis. \newpage
\section{Results and discussion}
\label{sec:Res}
The validation of the MRI measurement system is carried out through quantitative checks of structural integrity, mass flow rate, and divergence-free condition. Building on this validation, the flow behaviour within the three TPMS structures is analysed, with emphasis on characteristic flow patterns and vorticity. MRI measurements are used to validate the CFD simulations and to assess the consistency between the experimental and numerical results.
\subsection{Validation of structural integrity}
\label{sec:struc_inte}
The experimental MRI data are initially used to validate the additive manufacturing process by assessing the structural integrity of the printed modules using the signal intensity of the images. Visual inspection of the printed structures using the signal intensity of the images revealed no defects or anomalies in any of the modules. In addition to this qualitative assessment, a quantitative evaluation is performed by determining the cross-sectional area of each structure using MATLAB. This process consists of three steps: filtering of the data, binarisation of the signal intensity images, and comparison of the resulting cross-sectional area of the structures with the corresponding area obtained from the CAD files. \par
\begin{figure}[h!]
\centering
\includegraphics[width=0.6\linewidth]{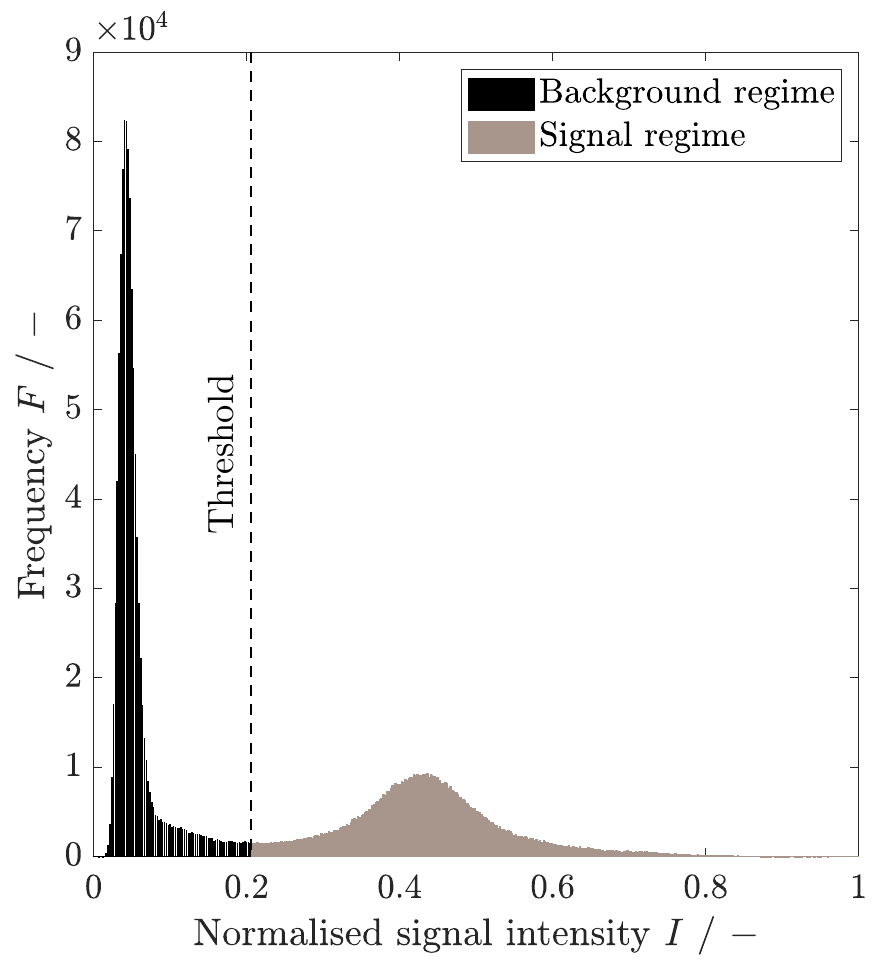}
\caption{Normalised signal intensity plotted against frequency for each slice, exemplarily shown for the Gyroid TPnS~$(\alpha = 0^\circ)$ at $Re_\text{S} = 156$ and $z=0~\text{mm}$ (see Tab.~\ref{tab:operating_cond}). The histogram of the signal intensity for a single slice shows two main peaks, with the local minimum used to filter and retain only the flow signal.} \label{fig:magnitude}
\end{figure}

The data are first limited to the flow-relevant region of the domain according to the procedure described by Bruschewski \textit{et al.}~\cite{bruschewski2016estimation, bruschewski2021magnetic}. In this approach, the normalised signal intensity is plotted against its frequency for each slice, as illustrated exemplarily in Fig.~\ref{fig:magnitude} for the Gyroid TPnS~$(\alpha=0^\circ)$ at $Re_\text{S}=156$ (see Tab.~\ref{tab:operating_cond}) and $z=0~\text{mm}$. In the complete analysis, the coordinate~$z$ denotes the axial position along the sample, where $z=0~\text{mm}$ corresponds to the first slice of the MRI acquisition.

\begin{figure*}[h!]
\centering
\includegraphics[width=1\linewidth]{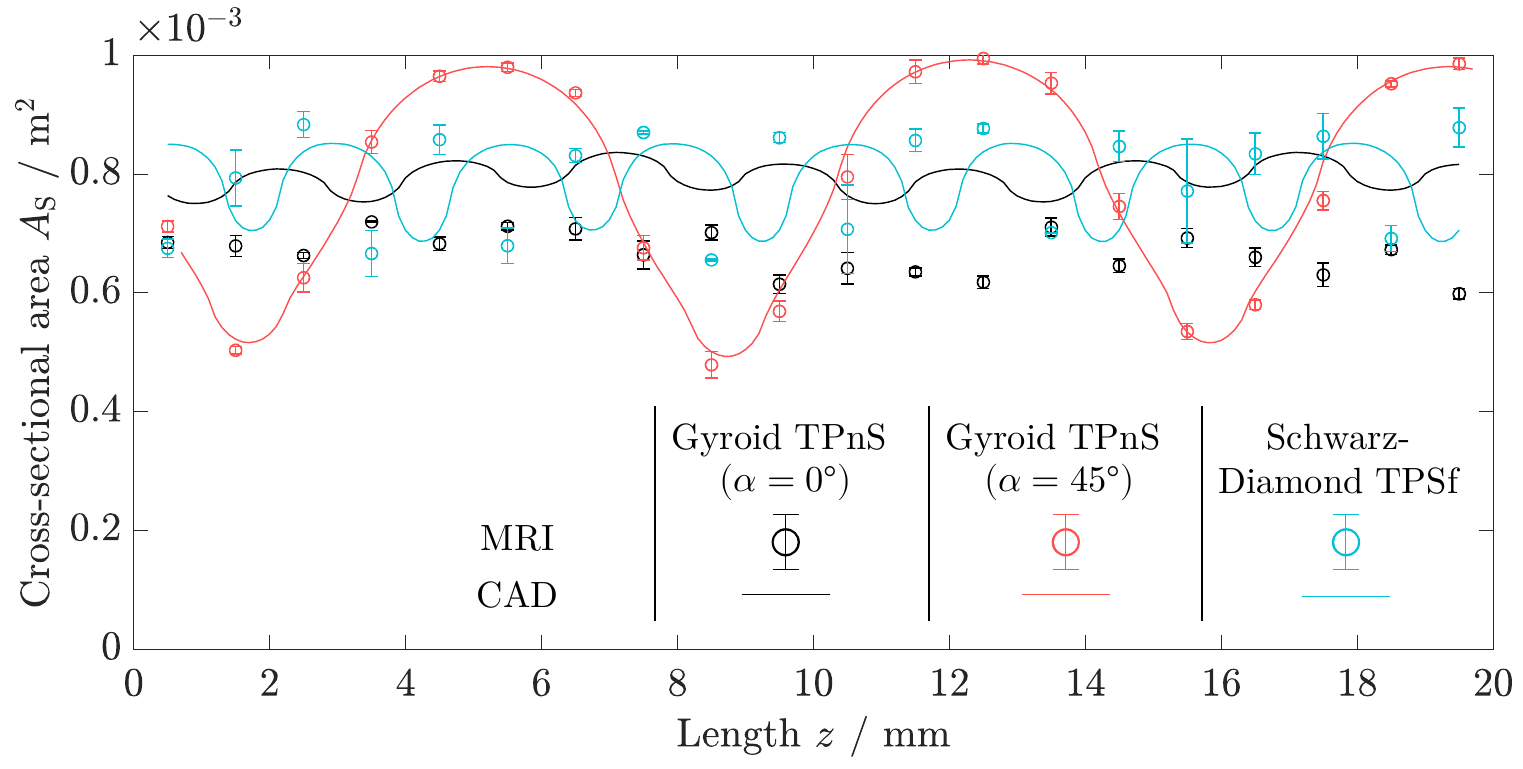}
\caption{Comparison of cross-sectional areas~$A_\text{S}$ derived from CAD and flow MRI data for the Gyroid TPnS structures $(\alpha = 0^\circ$ and $45^\circ)$ and the Schwarz-Diamond TPSf.}\label{fig:area}
\end{figure*}

Fig.~\ref{fig:magnitude} displays two prominent peaks: the lower-intensity peak corresponds to the background, encompassing the structure itself, motion artefacts, and artefact-free regions, whereas the higher-intensity peak represents the desired flow signal. The region between these peaks reflects partial volumes, i.e., voxels containing both fluid and solid material. Importantly, the partial volume region remains clearly distinguishable from the primary peaks. Given that both main peaks are larger than the partial volume contribution, it can be concluded that the spatial resolution is sufficiently high to enable an analysis of the structures~\cite{bruschewski2021magnetic}. Following the recommendation of Bruschewski \textit{et al.}~\cite{bruschewski2016estimation}, the local minimum for each slice between the two dominant peaks is determined and applied as a threshold. As a reference, the arithmetic mean threshold values for the Gyroid TPnS~$(\alpha = 0^\circ)$, Gyroid TPnS~$(\alpha = 45^\circ)$, and Schwarz-Diamond TPSf structures are $0.26 \pm 0.02$, $0.20 \pm 0.02$, and $0.13 \pm 0.01$, respectively. Each two-dimensional slice is binarised using the threshold determined for that slice. The processed MRI data are compared with the CAD data to evaluate the correspondence of the cross-sectional areas. The cross-sectional area is derived from the CAD data, which serve as the basis for both additive manufacturing and CFD simulations. For a quantitative assessment, the cross-sectional areas are determined from both the MRI data and the CAD models. The results along the axial direction for two unit cell lengths $20~\text{mm}$ are presented in Fig.~\ref{fig:area}.

From the analysis of Fig.~\ref{fig:area}, the Gyroid TPnS structures with $\alpha = 0^\circ$ and $\alpha = 45^\circ$ show mean deviations of 16~\% and 2.0~\% between the CAD-derived and MRI-derived data, respectively. The Schwarz-Diamond TPSf exhibits a deviation of 15~\% between the CAD-derived geometry and the MRI measurements, indicating that the datasets lie within the same order of magnitude and show overall agreement. The Gyroid TPnS~$(\alpha = 0^\circ)$ maintains an almost constant cross-sectional area along its length with variations smaller than 4~\%), which is reproduced by the MRI data. In contrast, the $\alpha = 45^\circ$ configuration shows a pronounced axial variation in the cross-sectional area, well captured by MRI. For the Schwarz-Diamond TPSf geometry, the general signal intensity images agree with the CAD model, but the detailed axial profile is more difficult to extract. This observation is attributed to the larger surface area of the structure (see Sec.~\ref{sec:AMLS}), which increases the fraction of wall-adjacent regions. In these regions, MRI measurements are characterised by a reduced signal-to-noise ratio due to partial-volume averaging, leading to a decreased detection accuracy. Minor registration errors cannot be excluded as a slight angular deviation of the experimental setup may have been present. Overall consistency between CAD and MRI can still be demonstrated. The additional optical inspection of the printed structures confirms that the additive manufacturing process is successful for all geometries and that the MRI data are valid. 

\subsection{Validation of velocity data}
\label{sec:velodata}
\begin{figure}[h!]
\centering
\includegraphics[width=0.6\linewidth]{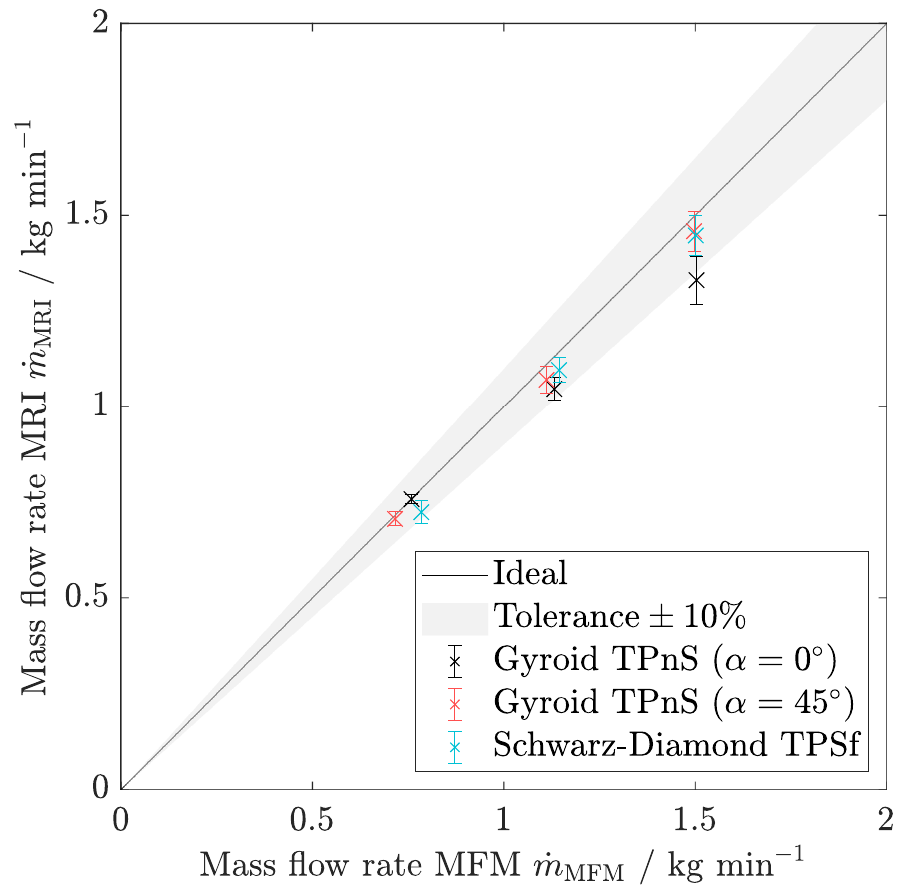}
\caption{Comparison of MRI-derived and measured mass flow rates of the Coriolis MFM, averaged across all slices.}\label{fig:mass-flow}
\end{figure}

For quantitative evaluation of flow data, the MRI velocity data are validated by means of mass flow rate and divergence. The mass flow rate is calculated according to Eq.~\ref{eq:mass-flow} across all slices and structures for the axial velocity. The comparison with measurements from the MFM is shown in Fig.~\ref{fig:mass-flow}. MRI-derived mass flow rates are averaged by arithmetic means over all slices. \par
On average, MRI underestimates the mass flow rate by 5.1~\% compared to the reference values obtained from the MFM. The deviation increases with increasing flow rate. The maximum relative error is observed for the Gyroid TPnS~$(\alpha = 0^\circ)$ at the highest flow rate, amounting to 12~\%, with the MRI consistently underestimating the mass flow. Such underestimation is well documented in the literature~\cite{o2008mri}. The observed deviations are small enough to consider the MRI measurements quantitatively reliable for analysing flow behaviour within the investigated TPMS structures. \par
The analysis of the mass flow rate considers only the axial velocity $w$. Therefore, the lateral velocities $u$ and $v$ are not directly validated with mass flow rate. To assess their accuracy, the divergence of the velocity field~$\boldsymbol{u}$ is computed according to Eq.~\ref{eq:div}. Subsequently, a mean divergence value is determined for each entire slice and operating point. In this paper, all values denoted by an overline represent arithmetic means. The values are in the order of magnitude $\overline{\text{div}\ \boldsymbol{u}} \lesssim 10^{-1}~\text{s}^{-1}$ for all structures and operating points. To relate these values to the overall mass flow rate, each mean divergence is multiplied by the fluid density~$\rho$ and the slice volume $V_\text{Slice} = \frac{\pi}{4} D^2 L_\text{Slice} \epsilon$ $(L_\text{Slice}=1~\text{mm})$, yielding the mass imbalance for the respective slice. Tab.~\ref{tab:divergence} summarises the mean mass imbalance~$\overline{\text{div}_\text{m}\ \boldsymbol{u}}$, averaged over all slices and pixel. \par
\begin{table*}[h!]%
\centering
\caption{Summary of the mean mass imbalance~$\overline{\text{div}_\text{m}\ \boldsymbol{u}}$ for all studied structures and operating points. The values are obtained by multiplying the mean slice divergence with the fluid density~$\rho$ and slice volume~$V_\text{Slice}$, providing a measure of the global consistency of the MRI velocity fields.}
\begin{tabular}{l c c c}
\hline Mass flow rate $\dot M\ /\ \text{g min}^{-1}$ & 755 & 1150 & 1500 \\ \hline
 \multicolumn{4}{c}{Mean mass imbalance $\overline{\text{div}_\text{m}\ \boldsymbol{u}}\ /\ \text{g min}^{-1}$} \\ \hline
Gyroid TPnS~$(\alpha = 0^\circ)$ & $-8.9 \pm 36.2$ & $-1.4 \pm 8.5$ & $-6.8 \pm 23.0$ \\ 
 Gyroid TPnS~$(\alpha = 45^\circ)$ & $ -4.3 \pm 12.6$ & $-0.33 \pm 20.1$ & $-5.4 \pm 30.3$ \\ 
 Schwarz-Diamond TPSf & $-1.7 \pm 19.9$ & $1.9 \pm 19.2$ & $-1.1 \pm 12.0$ \\ \hline
\end{tabular}
\label{tab:divergence}
\end{table*}
As presented in Tab.~\ref{tab:divergence}, the mean mass imbalance remains small relative to the measured mass flow rate, with a relative error below 6~\% for all cases. This confirms that the MRI-derived velocity fields are physically consistent, including the lateral velocity components. Furthermore, the divergence is independent of the operating conditions. 
The deviations are primarily caused by measurement noise. Furthermore, the slice thickness induces volumetric averaging effects, which intensify gradient variations in the MRI datasets, particularly in regions where the TPMS geometry varies within a single slice. Voxels intersecting solid boundaries can therefore produce artificially high local divergence values. Despite these effects, the global divergence analysis and its agreement with the measured mass flow rate confirm that the MRI velocity data are reliable and physically consistent, providing a reliable basis for further flow analysis.
\clearpage
\subsection{Flow dynamics within TPMS structures}
\label{sec:flow_dynamics}
\begin{figure*}[h!]
\centering
\includegraphics[page = 1, width=0.95\linewidth]{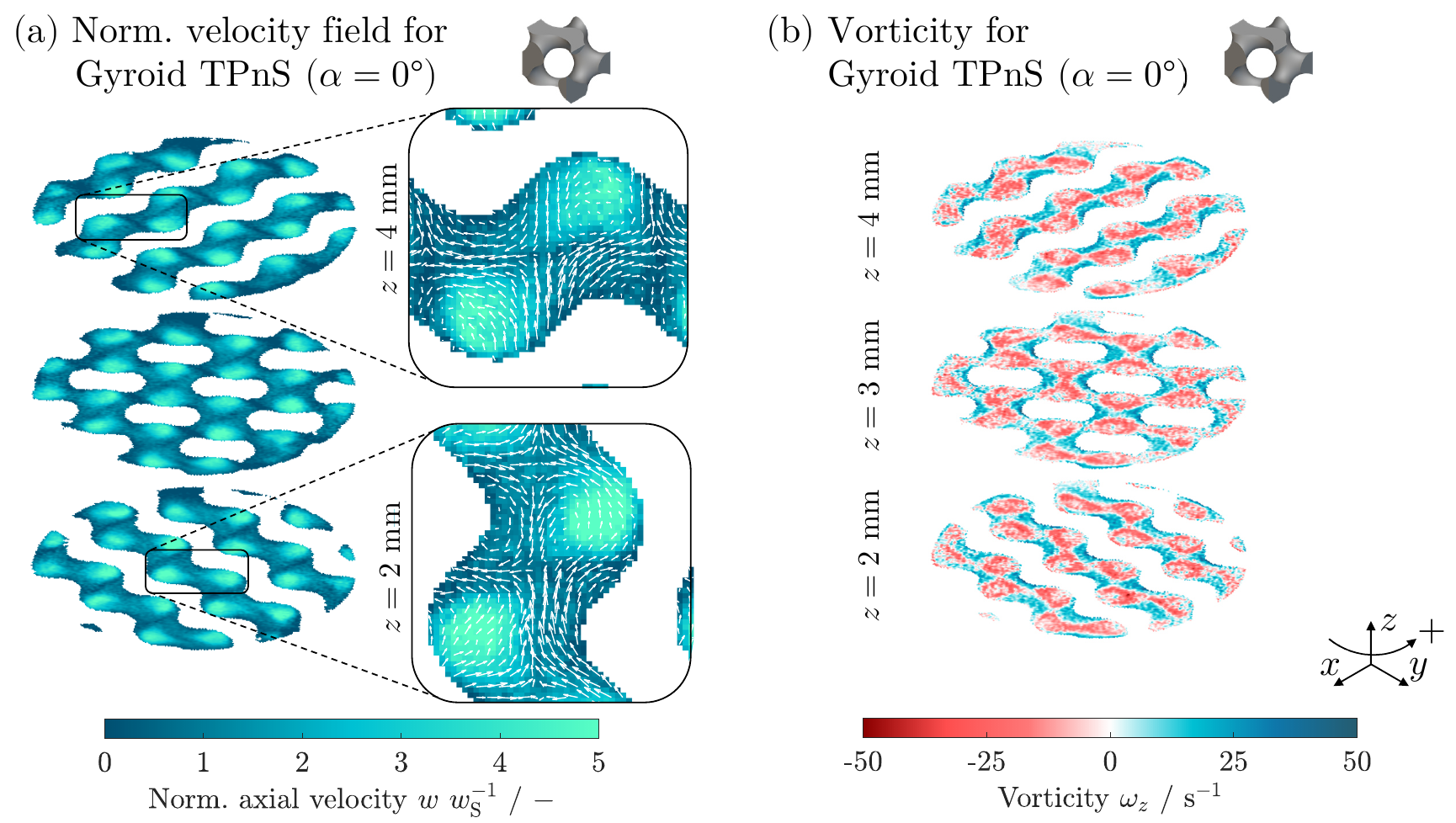}
\caption{MRI measurements are shown with in-plane velocity components $u$ and $v$ in the $x$- and $y$-directions are represented by arrows, while the colour scale indicates the axial velocity $w$, normalised by the superficial velocity $w_\text{S}$. Representative axial positions in (a). Corresponding vorticity fields~$\omega_z$ in (b), representing rotation around the $z$-axis and calculated according to Eq.~\ref{eq:vor} for Gyroid TPnS~$(\alpha = 0^\circ)$ at $Re_\text{S} = 156$.}\label{fig:2D_slice_G}
\end{figure*}

\begin{figure*}[h!]
\centering
\includegraphics[page = 1, width=0.95\linewidth]{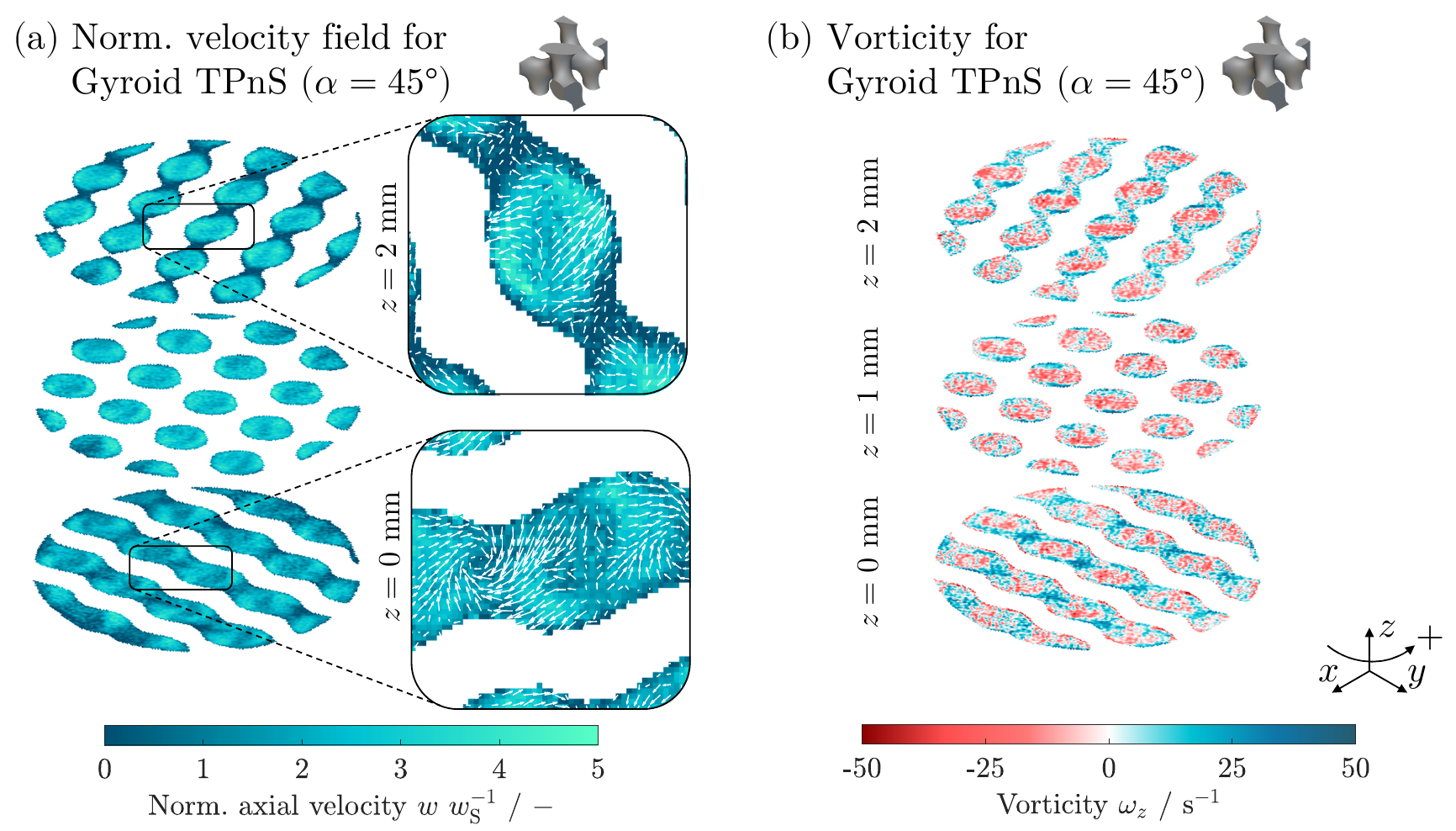}
\caption{MRI measurements are shown with in-plane velocity components $u$ and $v$ in the $x$- and $y$-directions are represented by arrows, while the colour scale indicates the axial velocity $w$, normalised by the superficial velocity $w_\text{S}$. Representative axial positions in (a). Corresponding vorticity fields~$\omega_z$ in (b), representing rotation around the $z$-axis and calculated according to Eq.~\ref{eq:vor} for Gyroid TPnS~$(\alpha = 45^\circ)$ at $Re_\text{S} = 155$.}\label{fig:2D_slice_G45}
\end{figure*}

\begin{figure*}[h!]
\centering
\includegraphics[page = 1, width=0.95\linewidth]{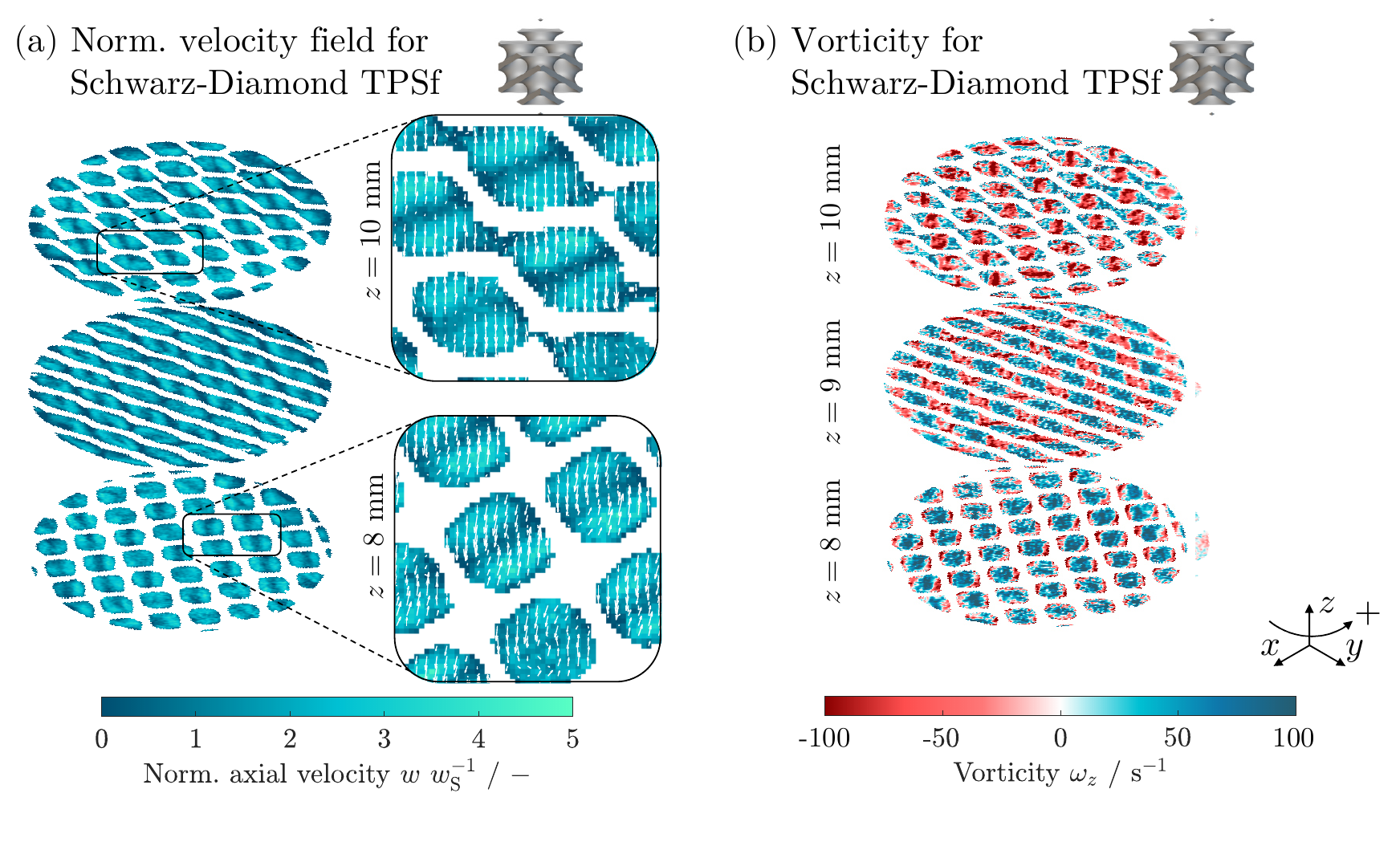}
\caption{MRI measurements are shown with in-plane velocity components $u$ and $v$ in the $x$- and $y$-directions are represented by arrows, while the colour scale indicates the axial velocity $w$, normalised by the superficial velocity $w_\text{S}$. Representative axial positions in (a). Corresponding vorticity fields~$\omega_z$ in (b), representing rotation around the $z$-axis and calculated according to Eq.~\ref{eq:vor} for Schwarz-Diamond TPSf at $Re_\text{S} = 129$.}\label{fig:2D_slice_SD}
\end{figure*}

Following the validation of both the structural information and the flow data, the analysis now focuses on the flow behaviour within the TPMS structures. The structures are compared at similar Reynolds numbers, with the Gyroid TPnS examined at a Reynolds number of $Re_\text{S}=156$ and the Schwarz-Diamond TPSf structure at $Re_\text{S}=129$. Based on these Reynolds numbers, the flow conditions fall within Darcy-Forchheimer regime, where laminar and steady behaviour without vortices is expected~\cite{dybbs1984new}. Figs.~\ref{fig:2D_slice_G}(a) to~\ref{fig:2D_slice_SD}(a) present three consecutive velocity fields at representative axial positions~$z$ measured by MRI velocimetry for the three TPMS structures. Arrows indicate the in-plane velocity components $u$ and $v$ in the $x$- and $y$-directions, while the colour scale represents the axial velocity $w$, normalised by the superficial fluid velocity $w_\text{S}$ (see Tab.~\ref{tab:operating_cond}). In Figs.~\ref{fig:2D_slice_G}(b) to~\ref{fig:2D_slice_SD}(b), the vorticity around the $z$-axis is shown, calculated according to Eq.~\ref{eq:vor}.

\clearpage
The Gyroid TPnS~$(\alpha = 0^\circ)$~(G) exhibits channelling behaviour, with axial velocities reaching five times the superficial velocity within the channels. This can be explained by the continuous channels running through the entire structure, generating high local channel velocities and a relative standard deviation of the axial velocity of $(87 \pm 6)~\%$, which indicates a heterogeneous flow behaviour.
Nevertheless, lateral flow structures are also observed. On average, the magnitudes of the lateral velocity components $u_\text{G}$ and $v_\text{G}$ correspond to $(38 \pm 4)~\%$ and $(45\pm 4)~\%$, respectively, of the mean axial velocity $\overline{w}_\text{G}$, indicating that the flow remains predominantly axial, despite noticeable lateral components. Characteristic helical flow structures arise as a direct consequence of the geometric features of the Gyroid TPnS~$(\alpha = 0^\circ)$. Between adjacent channels, counter-rotating vortices develop in the negative coordinate direction. The positive vorticity consistently appears between two neighbouring channels, promoting transverse exchange and thereby enhancing local mass transfer. This behaviour is clearly visible in Fig.~\ref{fig:2D_slice_G}(b), where positive vorticity values of $\overline{\omega}_{z\text{, G}}^+ = (12.3 \pm 0.6)~\text{s}^{-1}$ and negative vorticity values of $\overline{\omega}_{z\text{, G}}^- = (-13.5 \pm 0.7)~\text{s}^{-1}$ are observed. The nearly equal magnitudes of the positive and negative mean vorticities $\left(|\overline{\omega}_{z\text{, G}}^+| \approx |\overline{\omega}_{z\text{, G}}^-|\right)$ indicate the presence of symmetric, counter-rotating helical flow structures. Such flow patterns have also been reported in the literature and are in good agreement with the present findings~\cite{gado2024triply, padrao2024new}. However, although lateral velocity components are present, the overall flow is still dominated by pronounced channelling behaviour. Such a flow pattern is generally undesirable for chemical reactor applications, as it leads to non-uniform mass and heat transfer, potentially reducing reaction selectivity and overall efficiency. \par

In order to overcome these challenges, the rotated Gyroid TPnS structure is studied. The tilt misaligns the channels with the main flow direction, effectively reducing channelling while preserving the overall geometry and volume-specific surface area of the unit cell. In contrast to the unrotated configuration, the $45^\circ$-rotated Gyroid TPnS exhibits a different flow pattern, characterised by the absence of localised high-velocity regions. However, the overall level of flow heterogeneity remains comparable to the axial velocity standard deviation of $(84 \pm 26)~\%$. A closer inspection reveals that, in contrast to the unrotated Gyroid, the rotated configuration exhibits a pronounced axial dependence of the velocity standard deviation. For individual slices, velocity standard deviations of 43~\% are observed (e.g. at $z = 1~\text{mm}$ in Fig.~\ref{fig:2D_slice_G45}(a)), indicating a locally more homogeneous flow profile. In contrast, other axial positions exhibit higher values, reaching $76~\%$ and $105~\%$ at $z = 2~\text{mm}$ and $z = 3~\text{mm}$, respectively. Unlike the unrotated Gyroid TPnS, the high velocity standard deviations observed in the rotated configuration are associated with regions of low velocities, as illustrated by the close-up in Fig.~\ref{fig:2D_slice_G45}(a) at $z = 2~\text{mm}$. The observed fluctuations in the velocity distribution are associated with the strong axial variation of the free cross-sectional area along the structure (see Fig.~\ref{fig:area}). \par
The mean axial velocity reaches approximately $\overline{w}_\text{G45}\ w^{-1}_\text{S} = 1.5 \pm 0.4$, while the lateral velocity components account for $(49 \pm 9.4)~\%$ and $(52 \pm 12)~\%$ of the mean axial velocity in the $x$- and $y$-directions, respectively. This represents an increase compared to the unrotated case. As seen in Fig.~\ref{fig:2D_slice_G45}(a), at $z = 2~\text{mm}$, counter-rotating lateral flow structures emerge within the main channels. This behaviour is further reflected in the vorticity distribution shown in Fig.~\ref{fig:2D_slice_G45}(b), where alternating positive and negative vorticity regions are observed across the cross-section, with mean values for whole FOV of $\overline{\omega}_{z\text{, G45}}^+ = (12.2 \pm 1.0)~\text{s}^{-1}$ and $\overline{\omega}_{z\text{, G45}}^- = (-12.5 \pm 1.3)~\text{s}^{-1}$, indicative of symmetric, counter-rotating vortical motion. \clearpage
Overall, these changes indicate that unit-cell rotation redistributes the axial flow and promotes locally more homogeneous flow regions. A more homogeneous flow field is generally associated with enhanced heat and mass transfer~\cite{zhou2018reducing}. Accordingly, the Gyroid TPnS $(\alpha = 45^\circ)$ may promote improved heat and mass transfer performance by inducing locally more homogeneous flow conditions and a more uniform distribution of reactants. This effect arises from the suppression of channelling throughout the structure and the increase in the magnitude of lateral velocities, which facilitate enhanced transverse transport, despite comparable global flow heterogeneity. \par 

The Schwarz-Diamond TPSf~(SD) structure exhibits a larger volume-specific surface area and a smaller hydraulic diameter compared to the Gyroid TPnS (see Sec.~\ref{sec:Exsetup}). Consequently, for an equivalent Reynolds number, a higher mass flow rate is required within the structure. In the Schwarz-Diamond TPSf geometry, some isolated gas bubbles could not be avoided, as indicated by the region of reduced signal intensity on the right-hand side of the TPMS structure in Fig.~\ref{fig:2D_slice_SD}(a) at $z=9~\text{mm}$. However, their influence on the flow is expected to be negligible, since the bubbles are small compared to the channel cross-section and occur only irregularly. Fig.~\ref{fig:2D_slice_SD}(b) illustrates the corresponding flow field, revealing an even more homogeneous velocity distribution, with a mean normalised axial velocity at $\overline{w}_\text{SD}\ w^{-1}_\text{S}= 1.4\pm 0.2 $ across all slices. The fraction of axial velocities exceeding $3w_\text{S}$ serves as another indicator of flow homogenisation. For the Gyroid~TPnS~$(\alpha = 0^\circ)$, $(21 \pm 2)~\%$ of the axial velocities lie above this threshold, whereas the rotated configuration reduces this fraction to $(11 \pm 7)~\%$, and the Schwarz-Diamond~TPSf to only $(4.4\pm 1.3)~\%$. This progression reflects the increasingly uniform velocity distribution across the three geometries. \par

A closer examination of a single cell in the Schwarz-Diamond TPSf structure shows that the flow separates or merges in the centre due to locally reduced axial velocities. The velocity measurements further indicate that at $z = 8~\text{mm}$, the lateral velocity components are directed towards each other, whereas at $z = 10~\text{mm}$, they diverge. This alternating behaviour forms the characteristic merge-split flow pattern, which has been described in the literature as particularly beneficial for enhancing mixing and transport processes~\cite{gado2024triply}. As shown in Fig.~\ref{fig:2D_slice_SD}(b), the slice at $z=8~\text{mm}$ reveals that each individual channel exhibits a positive vorticity in the centre, while a negative vorticity forms at the channel boundaries. In the subsequent axial slice, the channels merge into an elongated longitudinal passage, giving rise to alternating regions of positive and negative vorticity. At $z=10~\text{mm}$ where the flow again separates into distinct channels, the vorticity pattern is reversed compared to the slice at $z=8~\text{mm}$. The Schwarz-Diamond TPSf exhibits vorticity magnitudes higher than those of the Gyroid TPnS structures, with mean values of $\overline{\omega}_{z\text{, SD}}^+ = (62.7 \pm 4.4)~\text{s}^{-1}$ and $\overline{\omega}_{z\text{, SD}}^- = (-65.1\pm 3.8)~\text{s}^{-1}$. This pronounced merge-split behaviour, together with elevated vorticity levels, is advantageous for chemical engineering applications, as it is indicative of enhanced cross-mixing and improved homogenisation of reactants.\par
\begin{figure*}[h!]
\centering
\includegraphics[width=1.0\linewidth]{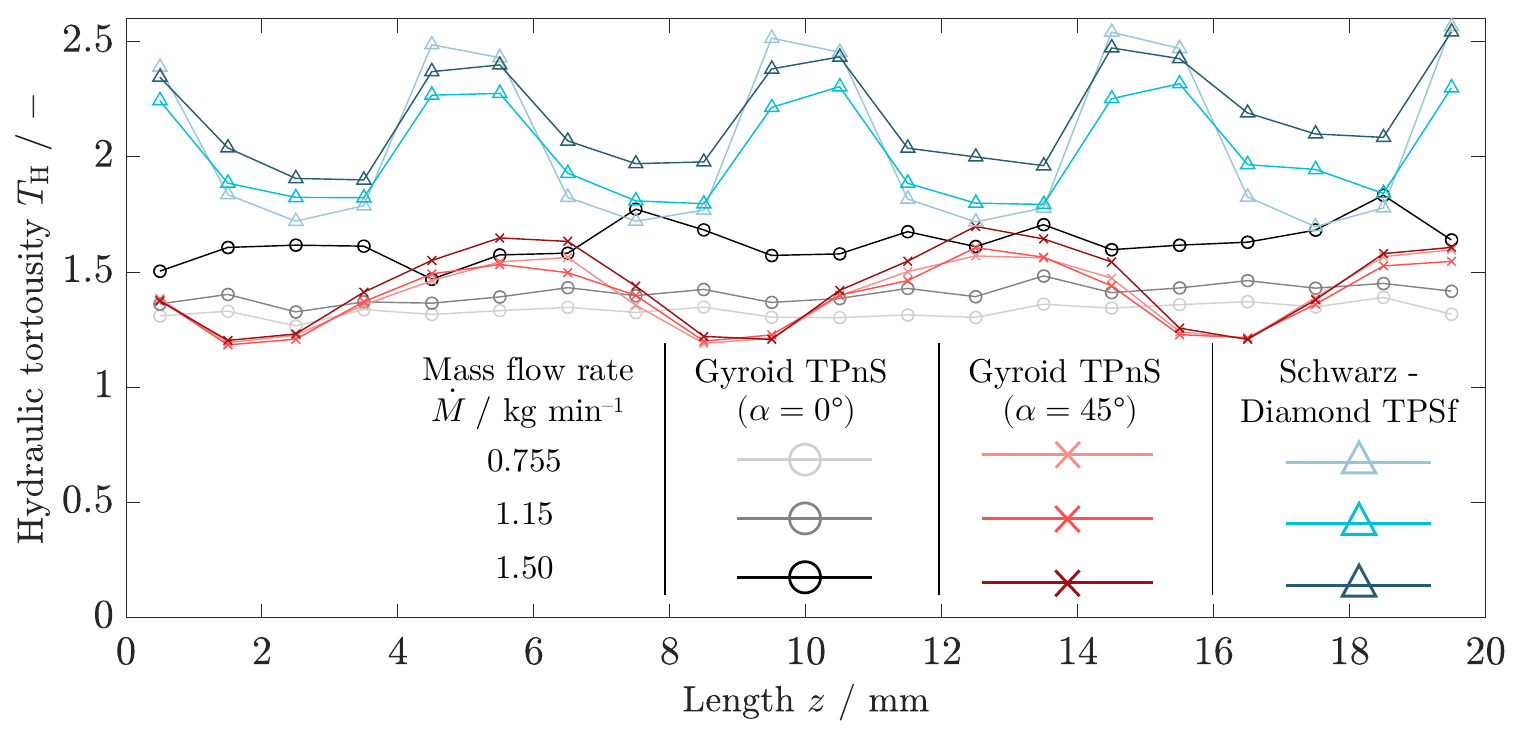}
\caption{Hydraulic tortuosity $T_\text{h}$ for the three TPMS structures (Gyroid TPnS, rotated Gyroid TPnS, and Schwarz-Diamond TPSf) at different Reynolds numbers based on MRI velocimetry. The tortuosity quantifies the ratio of radial to axial velocity components (see Eq.~\ref{eq:TH}), providing a measure of the degree of lateral flow and mixing intensity.}\label{fig:TH}
\end{figure*}
Following the analysis of local flow features, global flow quantities are evaluated. To further quantify the degree of lateral flow and mixing intensity, the hydraulic tortuosity, defined as the ratio of radial to axial velocity components and calculated according to Eq.~\ref{eq:TH}, is evaluated for all three TPMS geometries and operating conditions (Fig.~\ref{fig:TH}). The Schwarz-Diamond TPSf exhibits mean values 46~\% higher than those of the Gyroid TPnS structures, indicating a stronger cross-flow motion. For the Schwarz-Diamond TPSf, a mean hydraulic tortuosity of $\overline{T}_\text{h, SD}=2.10 \pm 0.25$ is obtained, whereas the Gyroid TPnS and rotated Gyroid TPnS show comparable magnitudes $\left (\overline{T}_\text{h, G}=1.46 \pm 0.053\ \text{and}\ \overline{T}_\text{h, G45}=1.42\pm 0.16 \right )$. The Gyroid TPnS reveals a moderate increase in hydraulic tortuosity with Reynolds number, rising by about 23~\% for a 100~\% increase in $Re_\text{S}$. Axial variations of the hydraulic tortuosity remain below 5~\%, reflecting the nearly constant cross-sectional area and strong influence of channelling throughout the whole structure. By contrast, the rotated Gyroid TPnS exhibits stronger axial fluctuations of 11~\%, consistent with its pronounced variation in free cross-sectional area (see Fig.~\ref{fig:area}). In comparison, the Schwarz-Diamond TPSf shows variations of hydraulic tortuosity of $12~\%$, which can be attributed to its non-continuous channels and intricate geometry that promote enhanced lateral mixing through merge-split patterns. Clarke \textit{et al.}~\cite{clarke2021investigation} reported $T_\text{h}=1.32$ for the Schwarz-Diamond TPSf structure under creeping and inertial flow conditions in smaller-scale systems. Compared to their findings, the present study shows by 59~\% higher hydraulic tortuosity values, consistent with the expected increase of lateral mixing at higher Reynolds numbers.
\clearpage

\subsection{Cross-validation of MRI and CFD results}
\label{sec:CFD_MRI}
The velocity fields obtained from MRI are compared with those from CFD simulations. CFD is employed here for cross-validation of the MRI data, while the MRI measurements simultaneously provide experimental validation of the numerically retrieved velocity fields. Although MRI and CFD data can be obtained for all three structures, the CFD-MRI comparison is performed on the Gyroid TPnS~$(\alpha = 0^\circ)$ structure, exemplary. This structure is selected because it exhibits pronounced channelling behaviour and the highest local velocities, representing the most demanding conditions for MRI flow measurements due to rapid spatial velocity variations. The Gyroid TPnS $(\alpha = 45^\circ)$ and Schwarz-Diamond TPSf structures exhibit more homogeneous flow behaviour, and the complete corresponding MRI are provided in~\cite{merbach2026data}. \par
The CFD data possess a 2.7-fold higher spatial resolution, highlighting the necessity of both methods. As explained in Sec.~\ref{sec:Comsetup}, the computational grid cannot be adapted to the lower spatial resolution of the MRI data due to the requirements of grid independence. At coarser grid resolutions, no accurate solution of the momentum equations can be obtained. The procedure used to align both datasets is described in Sec.~\ref{sec:Comsetup}. \par
\begin{figure}[h]
\centering
\includegraphics[page = 1, width=1.0\linewidth]{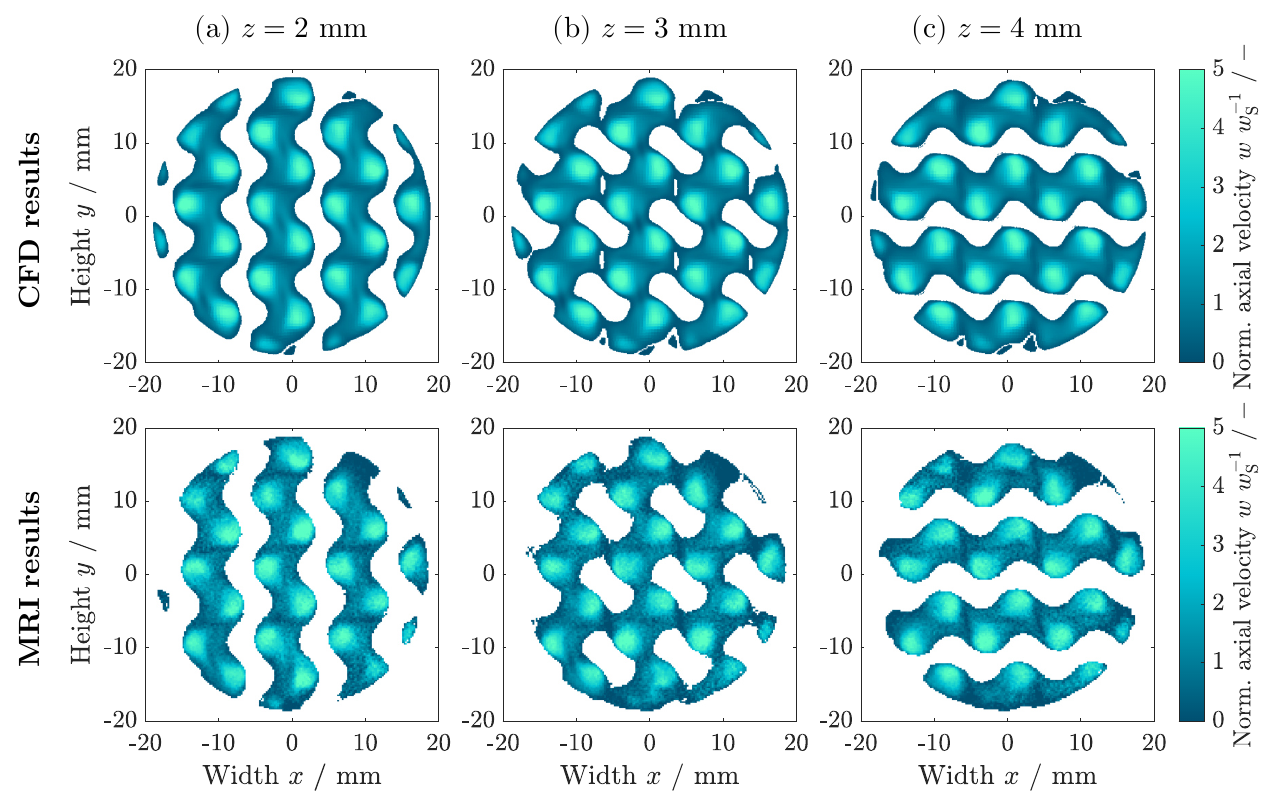}
\caption{Qualitative comparison of MRI and CFD velocity fields for Gyroid TPnS~$(\alpha = 0^\circ)$ at $Re_\text{S}=156$. Axial velocity fields are normalised by superficial velocity~$w_\text{S}$. The CFD data are volumetrically averaged over a slice thickness of 1~mm to match the MRI spatial resolution.}
\label{fig:CFD_MRI_qua_Re}
\end{figure}
\begin{figure}[h]
\centering
\includegraphics[width=0.6\linewidth]{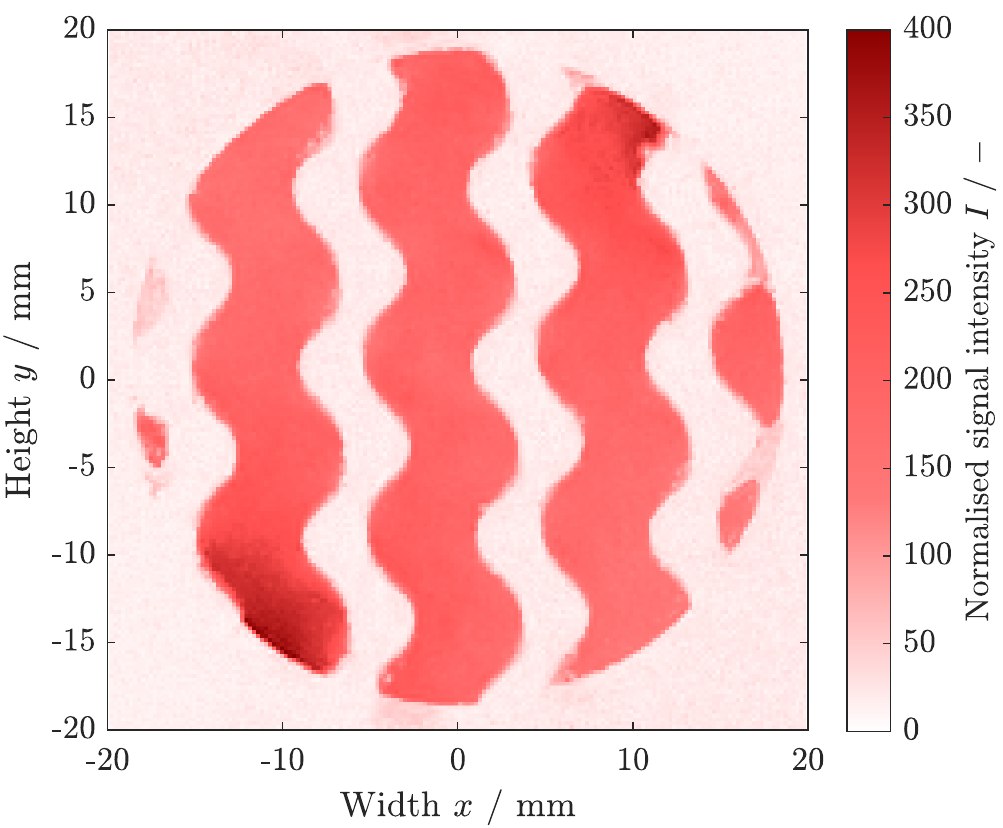}
\caption{Local MRI signal intensity distribution of the Gyroid TPnS ($\alpha = 0^\circ$) at $Re_\text{S}=156$, illustrating reduced signal intensity near the walls, particularly in regions farther from the RF coil, shown exemplarily at $z = 2~\text{mm}$.}\label{fig:A1_mag}
\end{figure}

A qualitative comparison of the axial velocity is shown in Fig.~\ref{fig:CFD_MRI_qua_Re} for three consecutive slice pairs of the Gyroid TPnS~$(\alpha = 0^\circ)$ at $Re_\text{S} = 156$. These data illustrate the close correspondence between the CFD and MRI results. The characteristic channels for the axial velocity of the Gyroid TPnS structures are clearly visible in both datasets, appearing at identical spatial locations and exhibiting comparable shapes and velocity magnitudes. Regions of low velocity, which are visible as horizontally aligned zones in the slice pair at $z=2~\text{mm}$, are likewise reproduced in both CFD and MRI results. Analysis of subsequent slice pairs confirms this consistency, indicating that both methods capture the same geometric and hydrodynamic features of the structure. Minor discrepancies occur near the wall regions, particularly within $(x, y) \in (15~\text{mm} \le x \le 20~\text{mm},\ -10~\text{mm} \le y \le 10$~mm; $-20~\text{mm} \le x \le -15~\text{mm},\ -10~\text{mm} \le y \le 10~\text{mm})$, where the MRI signal intensity is reduced due to the RF coil position at $(x, y) \in [-20~\text{mm}, -10~\text{mm}]$. This position results in high signal intensity along the cross-sectional diagonal, while regions further away from the RF coil exhibit weaker signals. The local signal intensity distribution shown in Fig.~\ref{fig:A1_mag} confirms this effect, indicating that the observed deviations originate from RF receive coil positioning rather than inaccuracies in the CFD data or the additive manufacturing process. Moreover, both MRI and CFD capture the three-dimensional velocity fields. To further illustrate this, a full three-dimensional visualisation of the lateral velocity components is provided in the Supplementary material. Characteristic minima and maxima of lateral velocities are identified at the same spatial locations in both datasets, demonstrating clear qualitative agreement between MRI and CFD. \par 

\begin{figure}[h]
\centering
\includegraphics[page = 1, width=1.0\linewidth]{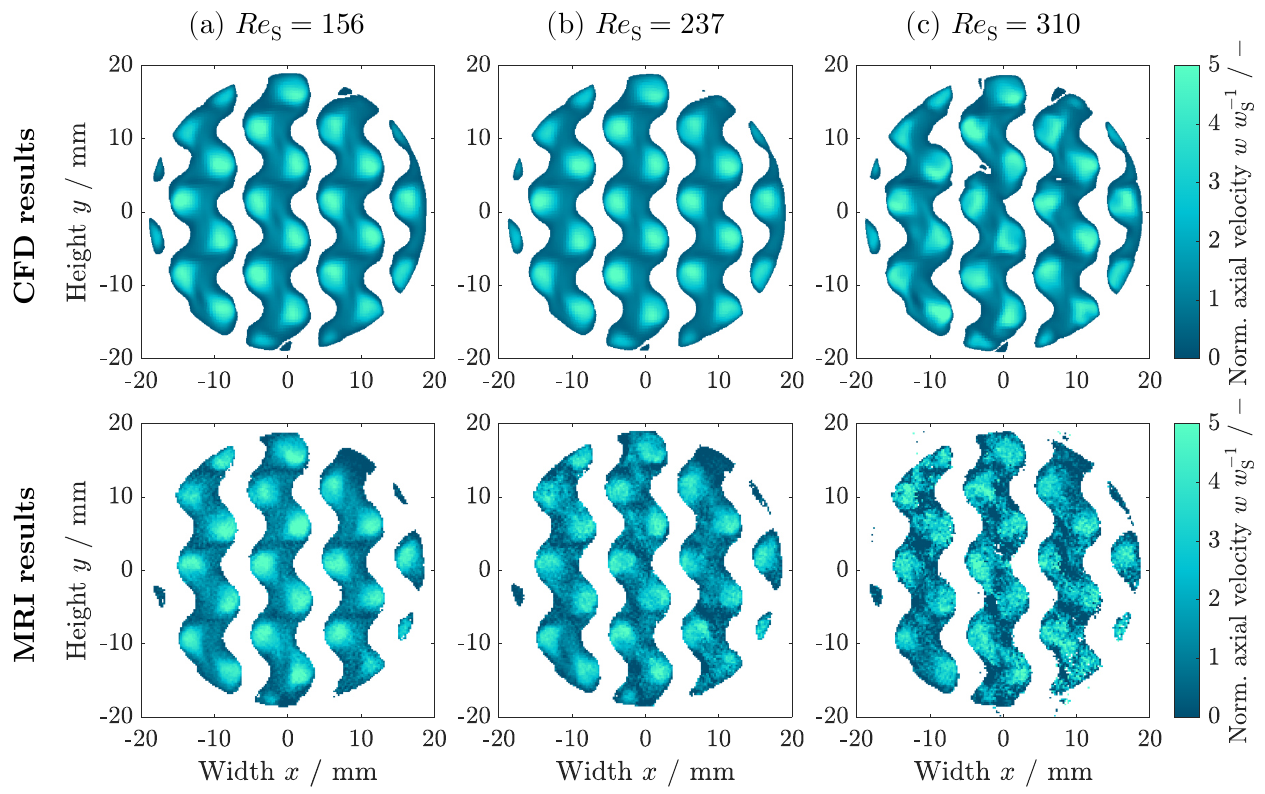}
\caption{Qualitative comparison of MRI- and CFD-derived velocity fields for the Gyroid TPnS ($\alpha = 0^\circ$) at $Re_\text{S}=156$ (a), $Re_\text{S}=237$ (b), and $Re_\text{S}=310$ (c) at $z=2~\text{mm}$. The axial velocity fields are normalised by the superficial velocity $w_\text{S}$.}\label{fig:MRI_CFD_Re}
\end{figure}

To assess the influence of the Reynolds number, additional consecutive slice pairs for all velocity components at the remaining operating conditions $Re_\text{S} = 237\ \text{and}\ Re_\text{S}=310$ are provided in the Supplementary material to facilitate a three-dimensional assessment of the velocity fields. As a representative example, Fig.~\ref{fig:MRI_CFD_Re} presents the slice at $z = 2~\text{mm}$ for the three operating points considered. \par
Overall, all datasets demonstrate strong agreement between CFD and MRI. However, the signal-to-noise ratio is observed to decrease with increasing volumetric flux, owing to the higher local velocities in the channels of the non-rotated Gyroid TPnS. The MRI data at $Re_\text{S} = 237$ remain of sufficient quality for reliable velocity estimation, while the signal-to-noise ratio decreases significantly at $Re_\text{S} = 310$ for all three velocity components. This behaviour may arise from the onset of vortex formation around $Re_\text{S} \approx 250$, resulting in a flow increasingly dominated by chaotic behaviour (transition from post-Forchheimer flow to turbulent flow)~\cite{dybbs1984new}. Flow turbulences and transient effects, which cause signal ghosts and motion artefacts in all phase-encoding directions, are also influencing the measurement unfavourably. The chaotic nature of the flow leads to heterogeneous velocities within individual voxels, causing additional signal attenuation~\cite{john2020magnetic}. The increasing turbulent flow behaviour is also observed in the CFD results: the flow field becomes increasingly irregular at this Reynolds number, indicating the onset of turbulence. \par

\begin{figure}[h!]
\centering
\includegraphics[width=1.0\linewidth]{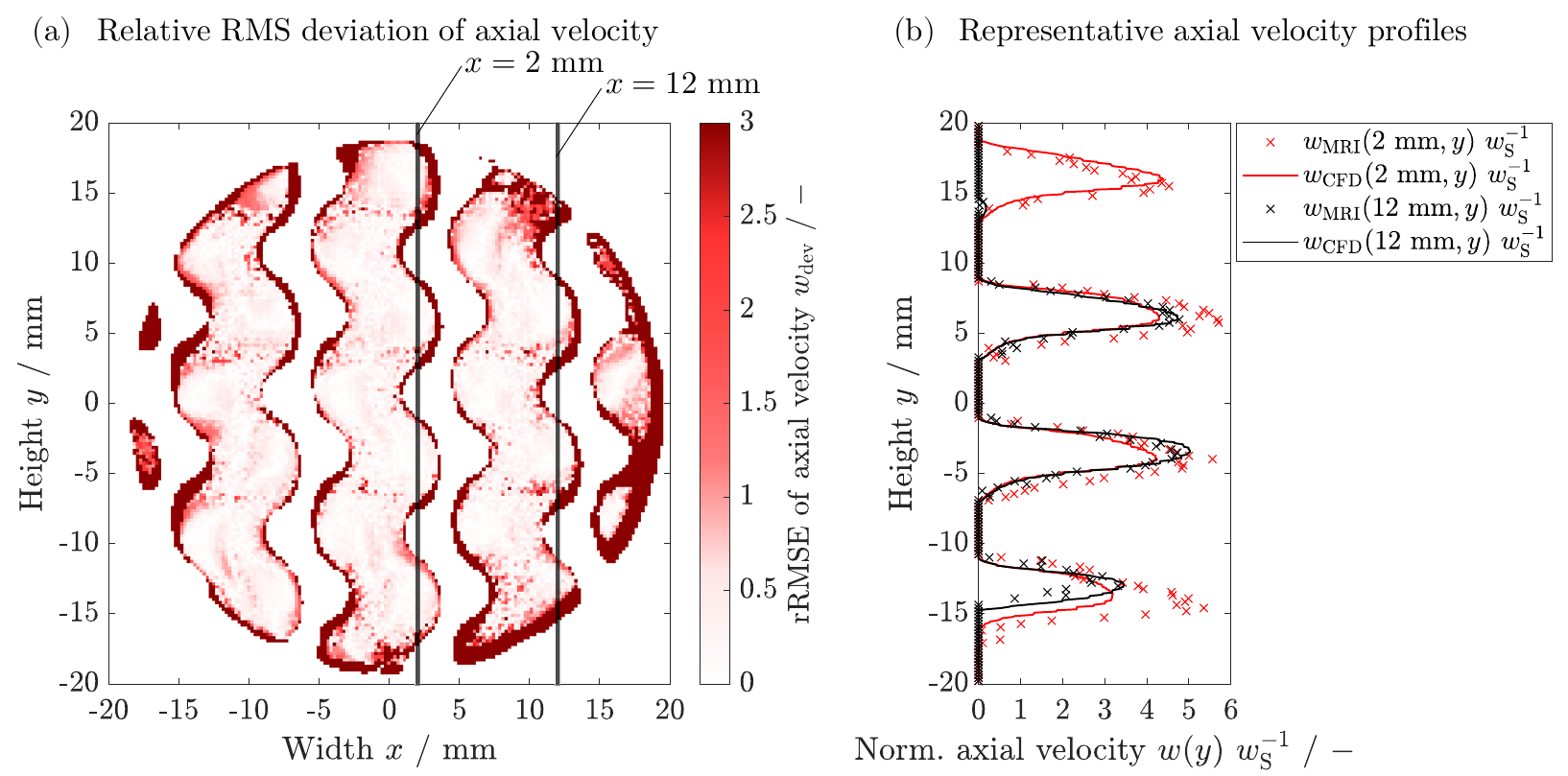}
\caption{Pixel-wise comparison of the axial velocity from CFD and MRI for the slice pair with Gyroid TPnS~$(\alpha = 0^\circ)$ at $z = 2~\text{mm}$ and $Re_\text{S} = 156$, showing the rRMSE deviation of the axial velocity in (a). In (b), two representative velocity profiles at fixed $x$-positions are presented.}\label{fig:CFD_MRI_qua}
\end{figure}

In addition to qualitative assessment, a quantitative comparison is essential to verify that the MRI not only reproduces the flow patterns but also accurately captures the corresponding velocity magnitudes. For this purpose, a pixel-wise comparison between the CFD and MRI datasets is performed. Since the MRI measurements feature a lower spatial resolution than the CFD data, the CFD velocity fields are convolved to match the MRI resolution prior to comparison. The relative Root Mean Square Error~(rRMSE) deviation of the axial velocity, $w_\text{dev}(x,y)$, is then evaluated for each in-plane location~$x,y$ according to
\begin{equation}
\label{eq_rRMSE}
w_\text{dev}(x,y) = \sqrt{\frac{\left ( w_\text{MRI}(x,y) - w_\text{CFD}(x,y) \right )^2}{(\overline{w}_\text{MRI, CFD})^2(x,y) }}\ .
\end{equation} 
Here, $w_\text{MRI}(x,y)$ and $w_\text{CFD}(x,y)$ denote the local axial velocities obtained from MRI and CFD, respectively, while $\overline{w}_\text{MRI,CFD}(x,y)$ represents their local mean value~\cite{clarke2021investigation, clarke2025investigation}. An example of this comparison is illustrated in Fig.~\ref{fig:CFD_MRI_qua}(a) for the MRI/ CFD slice pair $(z=2~\text{mm})$ shown in Fig.~\ref{fig:CFD_MRI_qua_Re}(a). An examination of Fig.~\ref{fig:CFD_MRI_qua}(a) shows that in regions where the CFD and MRI data overlap, mean rRMSE deviations of $\overline{w}_\text{dev}=40~\%$ are observed between the two datasets at $z =2~\text{mm}$. For $Re_\text{S} = 237$ and $Re_\text{S} = 310$, the deviations amount to $\overline{w}_\text{dev} = 49~\%$ and $\overline{w}_\text{dev} = 75~\%$, respectively (see Supplementary material for the corresponding quantitative comparisons). The mean rRMSE $\overline{w}_\text{dev}$ is calculated as the square root of the ratio of the separately summed numerator and denominator terms of Eq.~\ref{eq_rRMSE}. 
Reasons for the deviation are system-related imperfections, such as minor gas bubbles, local variations in the printing quality of additively manufactured structures, and residual registration inaccuracies between the MRI and CFD datasets. \par 
In addition, the decrease in signal-to-noise ratio with distance from the RF coil may contribute to the observed deviations. Additional sources of noise is another contributing factor. Near-wall regions, where partial-volume effects and signal loss lead to an under-representation of velocities within approximately~$500~\mum$ of the solid boundaries. Despite the good qualitative agreement, the quantitative discrepancies remain high, particularly at elevated Reynolds numbers, where the flow becomes increasingly chaotic, amplifying local velocity fluctuations and measurement uncertainties. Pixel-wise comparisons are more sensitive, as they capture local discrepancies at the resolution limit of the MRI system. In addition, small misalignments, local velocity fluctuations, and regions of low or near-zero flow may further amplify relative deviations, despite an accurate representation of the macroscopic flow features. \par
Thus, Fig.~\ref{fig:CFD_MRI_qua}(b) shows the normalised axial velocity profiles for two representative positions of constant width~$x$. The selected $x$-positions correspond to locations where transitions between structural elements and flow regions occur. The graph shows relative deviations of 10~\% and 14~\% at $x=2~\text{mm}$ and $x=12~\text{mm}$, respectively.
Relative deviations between CFD and MRI data are computed element-wise. Cases where both values are zero are set to 0~\%, while divisions by zero are treated as undefined. The mean relative deviation is then calculated using only values below the 95th percentile to mitigate the influence of outliers. These values indicate a strong agreement between CFD and MRI. Close to the pipe wall, detection becomes increasingly unreliable, leading to larger deviations. Consequently, the CFD simulations are cross-validated using direct comparison with the MRI measurements. Thus, CFD can act complementary to resolve fine-scale flow details in regions that are difficult to access experimentally, particularly near the walls and at higher Reynolds numbers. Despite these limitations, the overall agreement between CFD and MRI is strong, enabling cross-validation of both techniques.
\section{Conclusion and outlook}
\label{sec:Con}
The flow behaviour in additively manufactured TPMS structures is studied using a large-bore vertical 3~T MRI system. Cross-validation between the experimental data and the numerical simulations is performed. Three different TPMS structures are analysed under Darcy-Forchheimer regime. The experimental setup enables quantitative flow measurements in TPMS geometries with a diameter of $38~\text{mm}$ and lengths of up to $1000~\text{mm}$, thereby extending the accessible spatial scales compared to conventional horizontally orientated MRI systems. \par
Structural integrity is verified, and reliable velocity acquisition is ensured through mass-flow-rate validation with mean deviations of $5.1~\%$. Furthermore, the MRI system provides fully three-dimensional velocity fields with a relative mass divergence deviation below $6~\%$, demonstrating the physical consistency of the measurements. \par
The MRI measurements reveal that the studied structures exhibit different flow features. The unrotated Gyroid TPnS exhibits pronounced channelling due to continuous flow pathways, resulting in strong axial velocity dominance even though lateral velocity components reach approximately 42~\% of the axial mean. After rotating the unit cell, redistributes the axial flow and promotes locally more homogeneous flow regions, which is advantageous for chemical engineering applications. This demonstrates that future design of TPMS structures must account not only for the choice of unit cell but also for its orientation. The Schwarz-Diamond TPSf exhibits pronounced merge-split behaviour and shows the highest potential for chemical engineering applications, with a $46~\%$ increase in lateral mixing compared to the Gyroid TPnS structures. \par
Validation of numerical simulations and cross-validation of MRI data are performed. CFD velocity fields match MRI velocity fields quantitatively and qualitatively. Pixel-wise comparisons show good agreement in Darcy-Forchheimer regime with deviation of 40~\%. The resulting CFD-MRI agreement confirms the fidelity of the experimental measurements and the suitability of the CFD framework for resolving finer-scale flow structures. At higher Reynolds numbers, the increasingly irregular flow field indicates the onset of turbulence, requiring improved temporal resolution in MRI measurements as well as transient CFD simulations~\cite{clarke2025investigation}.\par
The combined experimental-numerical approach provides a foundation for future analysis of heat and mass transfer. By enabling spatially resolved characterisation of flow behaviour in complex TPMS reactor geometries, this work directly contributes to the SMART Reactors objective of reliably characterising transport processes in adaptive reactor systems. Future MRI developments are expected to enable the direct observation of chemical reactions, including the mapping of temperature and chemical composition within structured reactors~\cite{evans2006magnetic, rotzetter2020magnetic}. The flow-field characterisation presented forms a basis for these future studies. Beyond the present configuration, the vertical MRI system can accommodate larger experimental setups. With a bore diameter of $400~\text{mm}$, studies at industrially relevant scales become feasible, although further development of multi-channel detection systems will be required to maintain spatial and temporal resolution at larger column diameters.

\section{Declaration of Competing Interest}
The authors declare that they do not have any competing interests.

\section{Acknowledgements}
We thank Prof. Wilhelm Schwieger (FAU Erlangen-Nürnberg) for the valuable discussions on the results and for suggesting the idea of rotating the Gyroid TPnS unit cell. Publishing fees supported by Funding Program Open Access Publishing of Hamburg University of Technology (TUHH). The Magnetic Resonance Imaging System (large-bore 3~T vertical MRI system) was funded by the Deutsche Forschungsgemeinschaft (DFG, German Research Foundation) – project number 422037415 – under the “Major Research Instrumentation” funding program according to Art. 91b of the Basic Law (GG) and by The Free and Hanseatic City of Hamburg. This project is funded by the Deutsche Forschungsgemeinschaft (DFG, German Research Foundation) – SFB 1615 – 503850735. The authors gratefully acknowledge this funding.

\section{Author contributions}
Conceptualisation: TM, MA, CW, SIH, BB, AT, SA, CW, FK, MH, DH, IK, TK, AP, MS Experimental setup and measurements: TM, MA, SIH, BB, AT, SA Formal analysis and data curation: TM, MA, CW, BB, AT Writing-original draft: TM, MA, CW Writing-review and editing: TM, MA, CW, SIH, BB, AT, SA, CW, FK, MH, DH, IK, TK, AP, MS Funding acquisition: MH, DH, IK, TK, AP, MS All authors have read and agreed to the published version of the manuscript.

\section{Data availability}
The data that support the findings of this study are openly available at the following URL/DOI: https://doi.org/10.15480/882.16072~\cite{merbach2026data}.
\section{Declaration of generative AI and AI-assisted technologies in the writing process}
During the preparation of this work the authors used ChatGPT-4o in order to improve clarity and readability. The authors have thoroughly checked all the proposed edits and take full responsibility for the content of this manuscript.
\bibliographystyle{elsarticle-num} 
\bibliography{cas-refs.bib}
\appendix
\clearpage
\section{Nomenclature}
\begin{mdframed}
\subsection*{\textnormal{\textbf{Roman symbols}}}
\begin{tabularx}{\textwidth}{@{}lX@{}}
$A$ & Area / m$^2$ \\
$A_\text{P}$ & Pixel area / m$^2$ \\
$A_\text{S}$ & Cross-sectional area / m$^2$ \\
$A_\text{W, S}$ & Wetted surface area / m$^2$ \\
$a$ & Volume-specific surface area / $\text{m}^2\ \text{m}^{-3}$ \\
$c$ & Concentration / mol L$^{-1}$ \\
$D$ & Pipe diameter / m \\
$D_\text{h}$ & Hydraulic diameter / m \\
$d_{50}$ & Median particle diameter / m \\
div $\boldsymbol{u}$ & Divergence of the velocity field $\boldsymbol{u}$ / s$^{-1}$ \\
$\text{div}_\text{m}\ \boldsymbol{u}$ & Mass imbalance of the velocity field $\boldsymbol{u}$ / kg s$^{-1}$ \\
$F$ & Frequency / $-$ \\
$F_\text{S}$ & Safety factor / -- \\
$g$ & Gravitational acceleration / m s$^{-2}$ \\
$h$ & Average mesh size / m \\
$I$ & Signal intensity / -- \\
$L_\text{E}$ & Entrance length / m \\
$L_\text{E, M}$ & Entrance length of TPMS structures / m \\
$L_\text{M}$ & Module length / m \\
$L_\text{Slice}$ & Slice length / m \\
$\dot M$ & Mass flow rate / kg s$^{-1}$ \\
$n$ & Number of pixels / ${-}$ \\
$p$ & Pressure / Pa \\
$\Delta p$ & Pressure drop / Pa \\
$P$ & Convergence index / -- \\
$R^*$ & Grid resolution / -- \\
$r$ & Refinement factor / -- \\
$T$ & Temperature / K \\
$T_\text{h}$ & Hydraulic tortuosity / -- \\ 
$T_1$ & Longitudinal relaxation time / s \\
$V_\text{W}$ & Wetted volume / m$^3$ \\
\end{tabularx}
\end{mdframed} 
\clearpage
\begin{mdframed}
\subsection*{\textnormal{\textbf{Roman symbols continued}}}
\begin{tabularx}{\textwidth}{@{}lX@{}}
$\boldsymbol{u}$ & Vector of velocity field / m s$^{-1}$ \\
$u$ & Lateral velocity in $x$-direction / m s$^{-1}$ \\
$v$ & Lateral velocity in $y$-direction / m s$^{-1}$ \\
$w$ & Axial velocity / m s$^{-1}$ \\
$w_\text{dev}$ & Relative root mean square error deviation / -- \\ 
$w_\text{S}$ & Superficial velocity / $\text{m s}^{-1}$ \\
$\Delta x$ & Distance between nodes in $x$ / m \\
$\Delta y$ & Distance between nodes in $y$ / m \\
$\Delta z$ & Distance between nodes in $z$ / m \\
$x$ & Width / m \\
$y$ & Height / m \\
$z$ & Length / m \\
\end{tabularx}

\subsection*{\textnormal{\textbf{Greek symbols}}}
\begin{tabularx}{\textwidth}{@{}lX@{}}
$\alpha$ & Rotation angle / $^\circ$ \\
$\eta$ & Viscosity / Pa s \\
$\epsilon$ & Porosity / -- \\
$\varepsilon$ & Change of the numerical solution / -- \\
$\rho$ & Density / kg m$^{-3}$ \\
$\omega_z$ & Vorticity around $z$-axis / s$^{-1}$ \\
$\omega_z^+$ & Positive vorticity around $z$-axis / s$^{-1}$ \\
$\omega_z^-$ & Negative vorticity around $z$-axis / s$^{-1}$
\end{tabularx}

\subsection*{\textnormal{\textbf{Dimensionless numbers}}}
\begin{tabularx}{\textwidth}{@{}lX@{}}
$Re_\text{S}$ & Reynolds number for porous media 
\end{tabularx}
\end{mdframed} \clearpage
\begin{mdframed}
\subsection*{\textnormal{\textbf{Abbreviations}}}
\begin{tabularx}{\textwidth}{@{}lX@{}}
CAD & Computer-Aided Design \\
CFD & Computational Fluid Dynamics \\
ECT & Electrical Capacitance Tomography \\
ECVT & Electrical Capacitance Volume Tomography \\
FOV & Field Of View \\
G & Gyroid TPnS $(\alpha = 0^\circ)$ \\
G45 & Gyroid TPnS $(\alpha = 45^\circ)$ \\
GCI & Grid Convergence Index \\
M2D & Multi-two-Dimensional \\
MFM & Mass Flow Meter \\
MRI & Magnetic Resonance Imaging \\
PBF-LB/P & Laser-Based Powder Bed Fusion of Polymers\\
PIV & Particle Image Velocimetry \\
PTV & Particle Tracking Velocimetry \\
RF & Radio Frequency \\
rRMSE & relative Root Mean Square Error \\
SD & Schwarz-Diamond TPSf \\
SLA & Stereolithography \\
TPMS & Triply Periodic Minimal Surface \\
TPnS & Triply Periodic endo-Skeleton \\
TPSf & Triply Periodic Surface \\
VPP-UVL & Vat Photopolymerisation by Ultra Violet Laser beam\\

\end{tabularx}
\end{mdframed} 
\end{document}